\documentclass[sigconf]{acmart}

\usepackage{booktabs} 

\usepackage{tablefootnote}

\copyrightyear{2018} 
\acmYear{2018} 
\setcopyright{acmcopyright}
\acmConference[DLRS 2018]{3rd Workshop on Deep Learning for Recommender Systems}{October 6, 2018}{Vancouver, BC, Canada}
\acmBooktitle{3rd Workshop on Deep Learning for Recommender Systems (DLRS 2018), October 6, 2018, Vancouver, BC, Canada}
\acmPrice{15.00}
\acmDOI{10.1145/3270323.3270328}
\acmISBN{978-1-4503-6617-5/18/10}

\begin{document}
\title[News Session-Based Recommendations using Deep Neural Networks]{News Session-Based Recommendations using Deep Neural Networks}

\author{Gabriel de Souza Pereira Moreira}
\affiliation{%
  \institution{CI\&T}
  \city{Campinas}
  \state{SP}
  \country{Brazil}
}
\additionalaffiliation{%
	\institution{Brazilian Aeronautics Institute of Technology - ITA}
	\city{S\~ao Jos\'e dos Campos}
	\state{SP}
	\country{Brazil}
}
\email{gabrielpm@ciandt.com}

\author{Felipe Ferreira}
\affiliation{%
  \institution{Globo.com}
  \city{Rio de Janeiro}
  \state{RJ}
  \country{Brazil}
}
\email{felipe.ferreira@corp.globo.com}

\author{Adilson Marques da Cunha}
\affiliation{%
  \institution{Brazilian Aeronautics Institute of Technology - ITA}
  \city{S\~ao Jos\'e dos Campos}
  \state{SP}
  \country{Brazil}
}
\email{cunha@ita.br}

\renewcommand{\shortauthors}{DLRS 2018, October 02-07, 2018, Vancouver, Canada}

\begin{abstract}
News recommender systems are aimed to personalize users experiences and help them to discover relevant articles from a large and dynamic search space. Therefore, news domain is a challenging scenario for recommendations, due to its sparse user profiling, fast growing number of items, accelerated item's value decay, and users preferences dynamic shift. 

Some promising results have been recently achieved by the usage of Deep Learning techniques on Recommender Systems, specially for item's feature extraction and for session-based recommendations with Recurrent Neural Networks.

In this paper, it is proposed an instantiation of the CHAMELEON -- a Deep Learning Meta-Architecture for News Recommender Systems. This architecture is composed of two modules, the first responsible to learn news articles representations, based on their text and metadata, and the second module aimed to provide session-based recommendations using Recurrent Neural Networks.

The recommendation task addressed in this work is next-item prediction for users sessions: "what is the next most likely article a user might read in a session?" 

Users sessions context is leveraged by the architecture to provide additional information in such extreme cold-start scenario of news recommendation. Users' behavior and item features are both merged in an hybrid recommendation approach. 

A temporal offline evaluation method is also proposed as a complementary contribution, for a more realistic evaluation of such task, considering dynamic factors that affect global readership interests like popularity, recency, and seasonality.

Experiments with an extensive number of session-based recommendation methods were performed and the proposed instantiation of CHAMELEON meta-architecture obtained a significant relative improvement in top-n accuracy and ranking metrics (10\% on Hit Rate and 13\% on MRR) over the best benchmark methods.

\end{abstract}

%
%



\maketitle

\section{Introduction}
Recommender Systems (RS) have been increasingly popular in assisting users with their choices, thus enhancing their engagement and overall satisfaction with online services \cite{jawaheer2014modeling}. They are an important part of information and e-commerce systems, enabling users to filter through large information and product spaces.

Recommender systems have been researched and applied in online services from different domains, like music \cite{Bu2010} \cite{van2013deep} \cite{wang2014} (e.g., Spotify, Pandora, Last.fm), videos (e.g. YouTube \cite{davidson2010youtube}), people \cite{Badenes2014} (e.g., Facebook), jobs \cite{Bastian2014} (e.g., LinkedIn \cite{kenthapadi2017personalized}, Xing \cite{mishra2016bottom}), and research papers \cite{wang2011collaborative} \cite{Beel2013b} (e.g., Docear \cite{beel2013introducing}), among others.

\subsection{Deep Learning on Recommender Systems}

Deep Learning (DL) \cite{hinton2006} \cite{hinton2006b} \cite{bengio2007} \cite{bengio2009} is a hot topic in machine learning communities. The uptake of deep learning by RS community was relatively slow, as the topic became popular only in 2016, with the first Deep Learning for Recommender Systems workshop at the ACM RecSys 2016 \cite{hidasi2017dlrs}.

Early pioneer work applying used neural networks to RS was done in \cite{salakhutdinov2007}, where a two-layer Restricted Boltzmann Machine (RBM) slightly outperformed Matrix Factorization. 

After a winter on RS research using neural networks, Deep Collaborative Filtering was addressed by \cite{wang2015} and \cite{wu2016} using denoising auto-encoders \cite{vincent2008extracting}. Deep neural networks have recently been used to learn item features from unstructured data, like text \cite{bansal2016ask}, music \cite{van2013deep} \cite{wang2014}, and images \cite{mcauley2015image} \cite{he2016deep}.

Recurrent Neural Networks (RNN) possess several properties that make them attractive for sequence modeling of user sessions. In particular, they are capable of incorporating input from past consumption events, allowing to derive a wide range of sequence-to-sequence mappings \cite{donkers2017sequential}. After the seminal work of \cite{hidasi2016}, a research line has emerged on the usage of RNNs on session-based  \cite{hidasi2016parallel} \cite{hidasi2017} \cite{wu2016recurrent} \cite{liu2016context} \cite{smirnova2017contextual} and session-aware \cite{donkers2017sequential} \cite{quadrana2017personalizing} \cite{ruocco2017inter} recommendations. 

\subsection{News Recommender Systems}
Popular news portals, such as Google News \cite{das2007}, Yahoo! News \cite{Trevisiol2014}, The New York Times \cite{spangher2015}, Washington Post \cite{graff2015} \cite{bilton2016}, among others have gained increasing attention from a massive amount of online news readers.

Online news recommendations have been addressed by researchers in the last years, either using Content-Based Filtering \cite{li2011scene} \cite{capelle2012semantics} \cite{ren2013concert} \cite{ilievski2013personalized} \cite{mohallick2017exploring}, Collaborative Filtering \cite{das2007} \cite{diez2016}, and Hybrid approaches \cite{chu2009personalized}  \cite{liu2010personalized}  \cite{li2011scene} \cite{rao2013personalized} \cite{lin2014personalized} \cite{li2014modeling} \cite{trevisiol2014cold} \cite{epure2017recommending}.

News domain poses some challenges for Recommender Systems:

\begin{itemize}
    \item \textbf{Sparse user profiling} -- the majority of readers are anonymous and they actually read only a few stories from the entire repository. This results in extreme levels of sparsity in the user-item matrix, as users usually have tracked very little information about their past behaviour, if any \cite{li2011scene} \cite{lin2014personalized} \cite{diez2016};
    \item \textbf{Fast growing number of items} -- hundreds of new stories are added daily in news portals (e.g., over 300 in The New York Times \cite{spangher2015}). This intensifies the cold-start problem, as for fresh items you cannot count on lots of interactions before starting to recommend them \cite{diez2016}. For news aggregators, scalability problems may arise, as a high volume of news articles overload the web within limited time span \cite{mohallick2017exploring};
    \item \textbf{Accelerated decay of item's value} -- information value decays over time. This is specially true in the news domain, as most users are interested in fresh information. Thus, each item is expected to have a short shelf life \cite{das2007}; and 
    \item \textbf{Users preferences shift} - news topics of interest are not as stable as in the entertainment domain. Some user interests shift over time, while other long-term interests remain stable \cite{diez2016}. User's current interest in a session may be affected by his context (e.g., location, access time) \cite{diez2016} or by global context (e.g., breaking news or important events) \cite{epure2017recommending}.
\end{itemize}

\section{A Deep Learning Architecture for News Session-Based Recommendations}

In \cite{moreira2018chameleon}, it was proposed the CHAMELEON - a Deep Learning Meta-Architecture for News Recommender Systems. A meta-architecture is a reference architecture that collects together decisions relating to an architecture strategy. It might be instantiated as different architectures with similar characteristics that fulfill a common task, in this case, news recommendations.

\begin{figure}[h]
	\includegraphics[width=9cm]{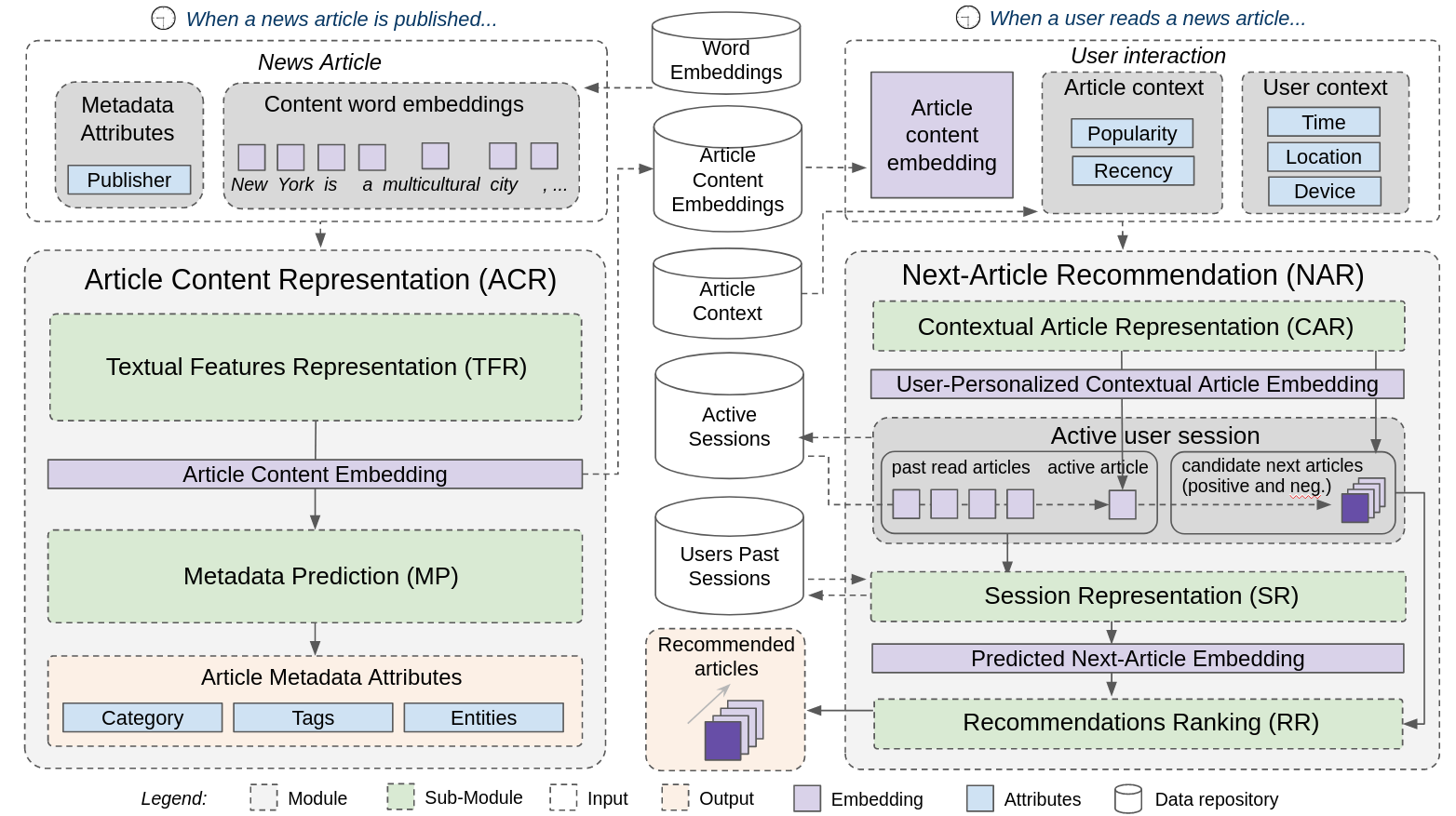}	\caption{\textit{CHAMELEON} - A Deep Learning Architecture for News Session-Based Recommendations} 
	\label{figure:chameleon}
\end{figure}

As shown in Figure~\ref{figure:chameleon}, CHAMELEON is composed of two complementary modules, with independent life cycles for training and inference: the Article Content Representation (\textit{ACR}) and the Next-Article Recommendation (\textit{NAR}) modules. 

\begin{figure}[h]
	\includegraphics[width=9cm]{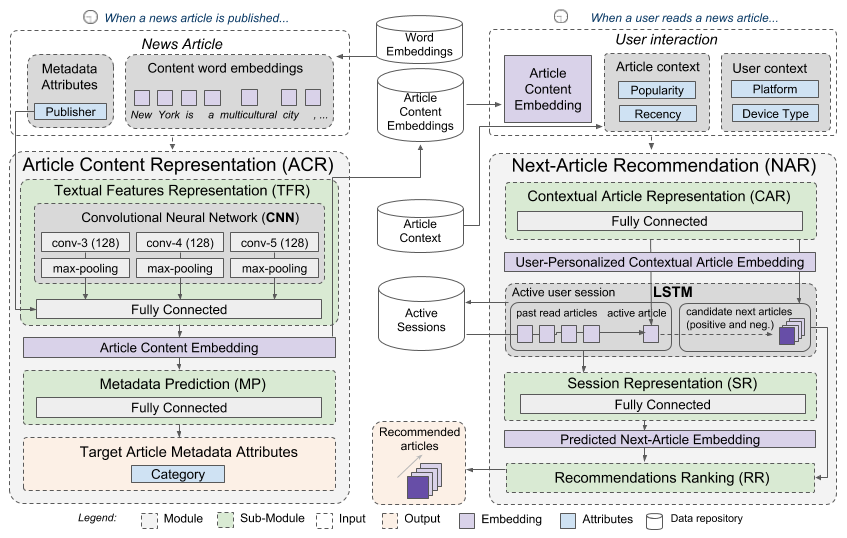}	\caption{An architecture instantiation of the CHAMELEON, using CNN and LSTM}
	\label{figure:chameleon_instantiation}
\end{figure}

In this work, the CHAMELEON meta-architecture was instantiated as a concrete architecture, presented in Figure~\ref{figure:chameleon_instantiation}. This instantiation of the \textit{ACR} module used Convolutional Neural Network (CNN) to learn textual features from news articles. In the \textit{NAR module}, the sequence of clicks from users sessions was modeled by Long Short-Term Memory (LSTM). 

The following sections describe this architecture instantiation of the CHAMELEON and also experiments performed on a large news dataset, compared to other session-based recommendation methods.

\subsection{Article Content Representation (ACR)}
The \textit{ACR} module is responsible to extract features from news articles text and metadata and to learn a distributed representations (embeddings) for each news article context. 

The inputs for the \textit{ACR} module are (1) article metadata attributes (e.g., publisher) and (2) article textual content, represented as a sequence of word embeddings. 

A common practice in Deep Natural Language Processing (NLP) it pre-training word embeddings using methods like Word2Vec \cite{mikolov2013} and GloVe \cite{pennington2014glove} in a larger text corpus of the target language (e.g., Wikipedia).

In this instantiation of the \textit{Textual Features Representation (TFR)} sub-module from  \textit{ACR} module, 1D CNNs was used to extract features from textual items, like in \cite{lee2016quote} and \cite{catherine2017transnets}.

Article's textual features and metadata inputs were combined by using a sequence of Fully Connected (FC) layers to produce  \textit{Article Content Embeddings}. 

For scalability reasons, \textit{Article Content Embeddings} are not directly trained for recommendation task, but for a side task of news metadata classification. In this work, they were trained to classify the category (editorial section) of news articles. 

In the last neural network layer, the \textit{softmax} function was used to normalize the output layer as a probability distribution, as follows:

\begin{equation} \label{eq:softmax}
\sigma(x_j) = \frac{e^{x_j}}{\sum_i e^{x_i}}.
\end{equation}

and cross-entropy log loss is used for optimization, as follows:

\begin{equation} \label{eq:crossentropyloss}
l(\theta) = -\frac{1}{N}(\sum_{i=1}^N y_i \cdot log(\hat{y_i})) + \lambda \lVert \theta \rVert_2,
\end{equation}

where  $ y $ is a vector with the one-hot encoded label for each instance, $ \hat{y} $ is the vector with the output probabilities for each class, previously normalized by \textit{softmax} function, $ \theta $, representing model parameters to be learned, and $ \lambda $ to control the importance of the regularization term, to avoid overfitting. 

The trained \textit{Article Content Embeddings} are stored in a repository, for further usage by \textit{NAR} module.

\subsection{Next-Article Recommendation (NAR)}

The \textit{Next-Article Recommendation (NAR)} module is responsible for providing news articles recommendations for active sessions.

Due to the high sparsity of users and their constant interests shift, this work leverages only session-based contextual information, ignoring possible users' past sessions. 

The inputs for the \textit{NAR} module are: (1) the pre-trained \textit{Article Content Embedding} of the last viewed article; (2) the contextual properties of the article (popularity and recency); and (3) the user context (e.g. time, location, and device). These inputs are combined by Fully Connected layers to produce a \textit{User-Personalized Contextual Article Embedding}, whose representations might differ for the same article, depending on the user context and on the current article context (popularity and recency).

The \textit{NAR} module uses a type of RNN -- the Long Short-Term Memory (LSTM) \cite{hochreiter1997long} -- to model the sequence of articles read by users in their sessions, represented by their \textit{User-Personalized Contextual Article Embeddings}. For each article of the sequence, the RNN outputs a \textit{Predicted Next-Article Embedding} -- the expected representation of a news content the user would like to read next in the active session.

In most deep learning architectures proposed for RS, the neural network outputs a vector whose dimension is the number of available items. Such approach may work for domains were the items number is more stable, like movies and books. Although, in the dynamic scenario of news recommendations, where thousands of news stories are added and removed daily, such approach could require full retrain of the network, as often as new articles are published.

For this reason, instead of using a softmax cross-entropy loss, the \textit{NAR} module is trained to maximize the similarity between the \textit{Predicted Next-Article Embedding} and the \textit{User-Personalized Contextual Article Embedding} corresponding to the next article actually read by the user in his session (positive sample), whilst minimizing its similarity with negative samples (articles not read by the user during the session).

With this strategy, a newly published article might be immediately recommended, as soon as its \textit{Article Content Embedding} is trained and added to a repository. The inspiration for this approach came from the DSSM \cite{huang2013learning} and its derived works for RS, like the MV-DNN \cite{elkahky2015multi}, the TDSSM \cite{song2016multi}, and the RA-DSSM \cite{kumar2017}, which uses a ranking loss based on embeddings similarity.

The \textit{Predicted Next-Article Embedding} and the \textit{User-Personalized Contextual Article Embedding}, further referred as $ p $ and $ item $, are vectors with the same arbitrary dimension. In Equation~\ref{eq:relevance}, it is defined a relevance function $ R $ as the cosine similarity between $ item $ and $ p $, as shown in Equation~\ref{eq:cosine_sim}.

\begin{equation} \label{eq:relevance}
R(s, item) = cos(s, item)
\end{equation}

\begin{equation} \label{eq:cosine_sim}
cos(\theta) = \frac{a \cdot b}{\lVert a \rVert \lVert b \rVert }
\end{equation}

Ranking-based loss functions are usually suitable for Top-N recommendations. The objective of the \textit{NAR} module is to produce a ranked list of the next likely article ($ item \in  D $, where $ D $ is the set of all items) the user will read in the session (next-click prediction). Thus, the model should learn to maximize the similarity between the \textit{Predicted Next-Article Embedding} ($ p $) and the \textit{User-Personalized Contextual Article Embedding} of the next article read by the user ($ item^+ $), whilst minimizing the pairwise similarity between $ p $ and \textit{User-Personalized Contextual Article Embeddings} of negative samples ($ item^- \in D^- $), where $ D^-  $ is the set of all items not read by the user in his session. 

As $ D $ set may be a very large in news domain, it is approximated as a set $ D' $ -- the union of the unit set with the clicked item (positive sample) $ \{item^+\} $ and a set with random negative samples from $ D^- $. 

The posterior probability of the next-clicked article given an active user session was computed by using a \textit{softmax} function over the relevance score proposed by the DSSM \cite{huang2013learning}, as shown in Equation \ref{eq:prob_click},

\begin{equation} \label{eq:prob_click}
P(item+ \mid s) = \frac{ exp(\gamma R(p, item+)}{\sum_{\forall item \in D'}{ exp(\gamma R(p, item)}}
\end{equation}

where $ \gamma $ is a smoothing factor (usually referred to as \textit{temperature}) for the softmax function, which may be a trainable parameter or empirically set on a held-out data set.

The \textit{NAR} module neural network parameters are estimated to maximize the likelihood of the next-clicked article given the user session. The loss function to be minimized, also introduced by the DSSM \cite{huang2013learning}, is shown in Equation \ref{eq:loss},

\begin{equation} \label{eq:loss}
l(\theta) = -log \prod_{(p,item^+)}{P(item^+ \mid s)},
\end{equation}

where $ \theta $ represent the model parameters to be learned. Since $ l(\theta) $ is differentiable w.r.t. to $ \theta $, the \textit{NAR} module is trained using back-propagation on gradient-based numerical optimization algorithms.

\section{Experiments}

The proposed instantiation of CHAMELEON meta-architecture was implemented using TensorFlow \cite{abadi2016tensorflow}, a popular Deep Learning framework. The source code for this neural architecture and also the baseline methods implemented for these experiments was open-sourced \footnote{https://github.com/gabrielspmoreira/chameleon\_recsys}.

The neural network models were trained and evaluated on Google Cloud Platform ML Engine, a managed platform for Deep Learning. 

For these experiments, a proprietary dataset was provided by the Globo.com\footnote{http://g1.globo.com/}. The Globo.com is the most popular news portal in Brazil, with more than 80M unique users and 100k new contents per month. The dataset sample contained user interactions from Oct. 1 to 16, 2017, including more than 3 million clicks, distributed in 1.2 million sessions from 330,000 users who read more than 50,000 different news articles during that period.

\subsection{Training and evaluation of the ACR module}
\begin{figure}[h]
	\includegraphics[width=7cm]{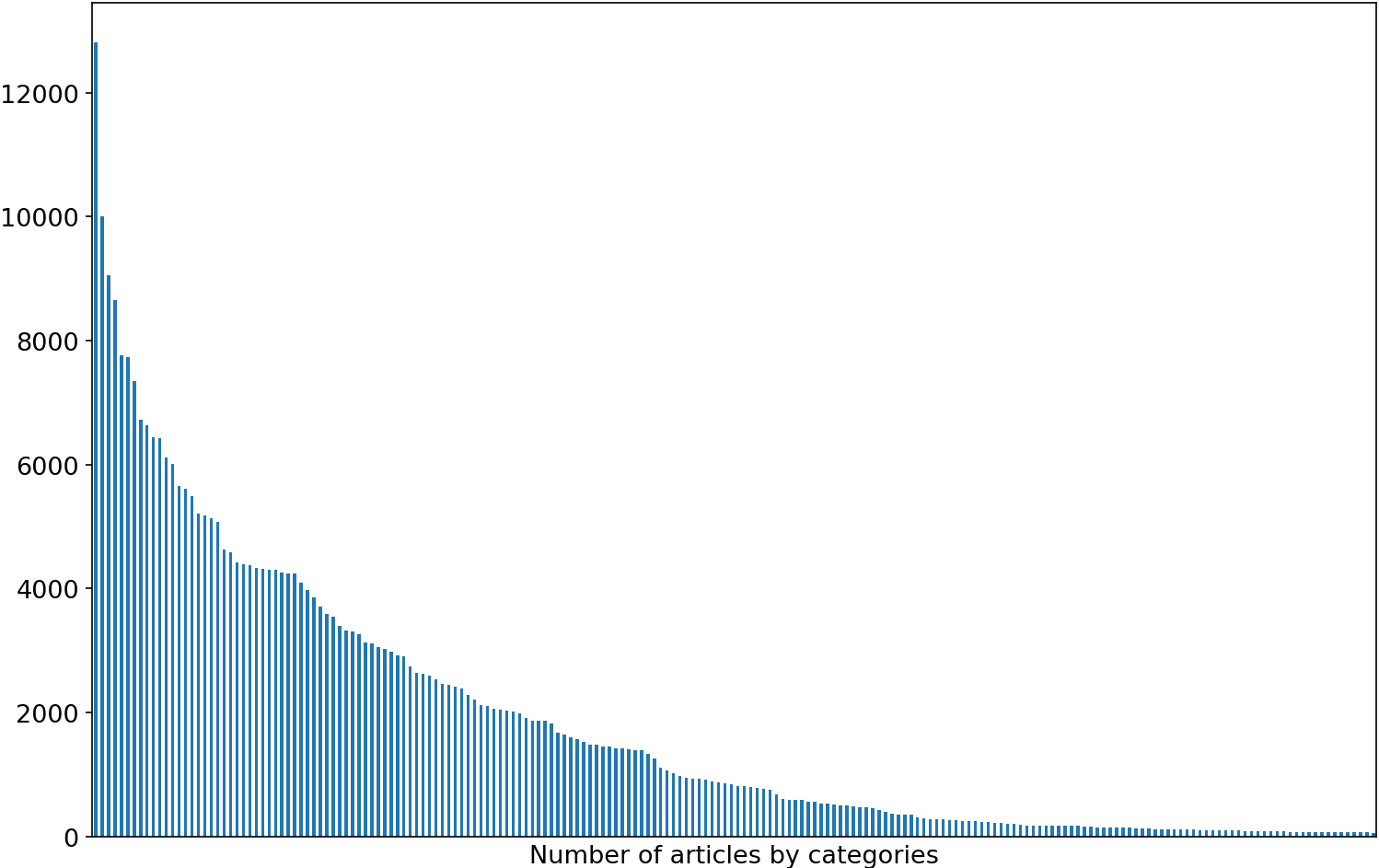}	\caption{Distribution of number of articles by categories}
	\label{figure:categories_count}
\end{figure}

The \textit{ACR} module was used to learn the \textit{Article Content Embeddings} for news articles. The model was trained to classify articles categories (editorial subsection in the news portal) based on its textual content and metadata. The dataset contained 364k news articles from 461 categories. The distribution of the top 200 categories can be seen in Figure~\ref{figure:categories_count}.

Articles texts were represented by sequences of pre-trained word embeddings for Portuguese language \footnote{It was used a pre-trained Word2Vec \textit{skip-gram} model (300-dims) for Portuguese language, available in http://nilc.icmc.usp.br/embeddings}.
Textual representation was learned by three 1D CNN layers, with window sizes of 3, 4, and 5 (to model word n-grams), each with 128 filters. After max-pooling operation, each of the three layers outputs feature maps (128-dims vectors), which were concatenated with other article's metadata (publisher) by means of a Fully Connected layer.

Training and evaluation were performed using the same dataset, as the objective for this network was not generalization, but to learn representations (\textit{Article Content Embeddings}) for articles content and metadata, with dimension size of 250.

\begin{figure}[h]
	\includegraphics[width=7cm]{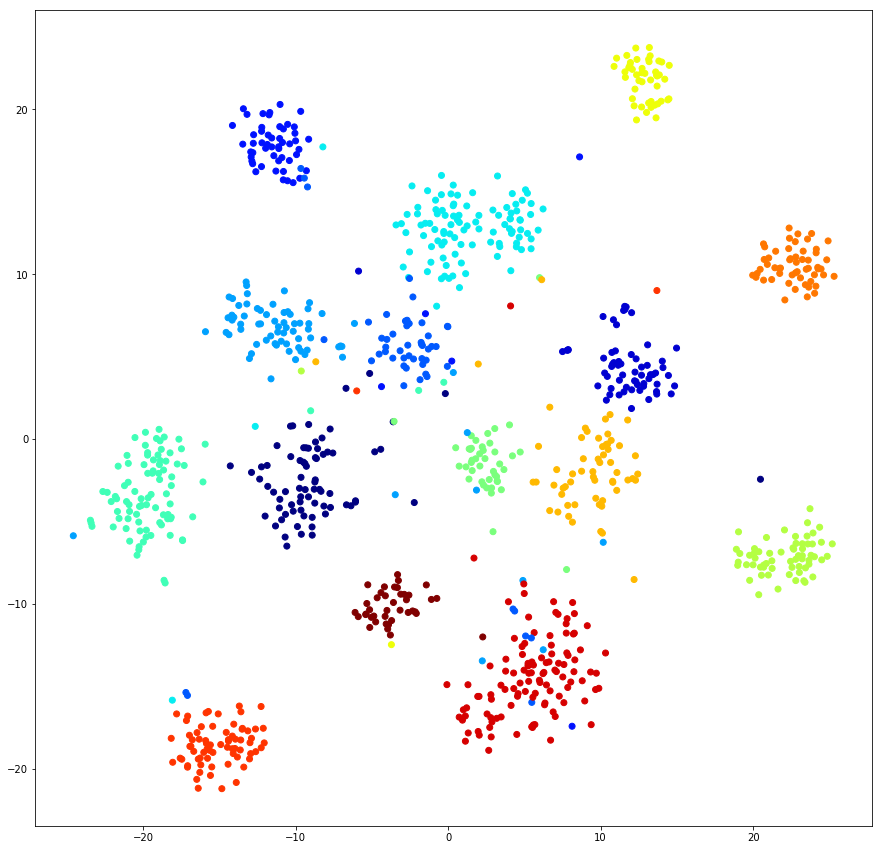}	\caption{t-SNE visualization of sampled \textit{Article Content Embeddings}, for the top 15 categories}
	\label{figure:acr_embeddings}
\end{figure}

Figure \ref{figure:acr_embeddings} presents a visualization produced using t-SNE, with sampled \textit{Article Content Embeddings} for the 15 categories with most articles. It can be observed that articles are clustered around their categories, which is expected, since the embeddings were trained to classify articles categories.

After the training, \textit{Article Content Embeddings} were persisted in a repository, for further usage by \textit{NAR} module. 

\subsection{Training and evaluation of the NAR module}
In the Globo.com dataset, a session represents a sequence of user clicks with no more than 30 minutes between interactions.

To train \textit{NAR} module, user interactions sequences were grouped by session and ordered by event time. Sessions with only 1 interaction (invalid for next-click prediction) and with more than 20 interactions (outlier users, or possibly bots) were discarded.

Each training or evaluation mini-batch was composed by a number of sessions (sequences of clicked items), padded to the largest session within the mini-batch.

A cross-validation approach is commonly used to evaluate machine learning models, where samples are split randomly to train or validation sets. Therefore, ignoring the temporal factor of news interactions would be unrealistic. 

News readership is very dynamic, as global interests may suddenly shift due to breaking events (e.g., natural disasters, or a royal family member birth) or may follow some seasonality (e.g., soccer during the World Cup, politics during a presidential election, etc.) \cite{chu2009personalized} \cite{gulla2016intricacies}. The popularity of a news article often changes very quickly, decaying in function of hours.

For this reason, it was devised for \textit{NAR} module a temporal offline evaluation method, which emulates a real-world scenario of continuously training the model with streaming user clicks and deploying a new trained model once an hour, to provide recommendations for new user sessions. The temporal offline evaluation method is described as follows: 

\begin{enumerate}
  \item Train the \textit{NAR} module with sessions within the active hour; and
  \item Evaluate the \textit{NAR} module with sessions within the next hour, for the task of the next-click prediction.
\end{enumerate}

This method is also scalable because, as each session is used only once for \textit{NAR} module training (online learning), there is no need to train on past sessions again (full retrain).

During the training and evaluation loop, in order to keep the consistency of temporal contextual information of articles (popularity and recency), the following method was created:

\begin{enumerate}
  \item Keep a global buffer with the last N clicks (article reads), considering all users;
  \item Compute articles recent popularity by counting their clicks within the buffer; and
  \item Compute articles recency as the number of elapsed hours since article was published; and
  \item For each article read by a user, look up for the updated article context features (recent popularity and recency).
\end{enumerate}

For each clicked article, the features for the \textit{NAR} module was its pre-trained \textit{Article Content Embedding}; the article context attributes (popularity and recency), smoothed by a \textit{log} function; and the user context attributes: platform (web, app) and device type (desktop, mobile, tablet). These inputs were concatenated and normalized with Layer Normalization technique \cite{ba2016layer}.

The \textit{NAR} module requires a number of negative samples to perform training and evaluation. The strategy adopted was to consider as negative samples any article not read within the session, which was read in other sessions from the training/evaluation mini-batch, as proposed in \cite{hidasi2016}. When there are not enough negative samples within the batch, items are uniformly sampled from a global buffer with the last N clicks. Such approach may help the network to learn to distinguish between the next clicked item (positive item) and other strong negative candidates (articles recently clicked by other users). 

For training, 7 negative samples were used for each session, and for evaluation, 50 negative samples. In this setting, each recommender algorithms is expected to rank the set composed by the clicked item and the negative samples.

The Top-N evaluation metrics used in this study were Hit Rate (HR@5) \cite{ludewig2018evaluation} which checks whether the clicked item is present in the top-5 ranked items, and Mean Reciprocal Rank (MRR@5) \cite{hidasi2016} \cite{jugovac2018streamingrec} \cite{ludewig2018evaluation}, a ranking metric, sensitive to the position of clicked item, which assigns higher score at top ranks.

Hyperparameters were tuned for \textit{NAR} module using in a separate period of the same dataset. The LSTM layer had 255 units and \textit{Predicted Next-Article Embeddings} dimension size was 1024. The best training settings were a mini-batch size of 256 sessions, with a learning rate of 1e-3, a $ L_2 $ regularization factor of 1e-4 and Adam \cite{kingma2014adam} as a gradient based optimizer to learn model parameters. Xavier initialization method \cite{glorot2010understanding} was used to initialize model parameters, as hyperbolic tangent (\textit{tanh}) was the mostly used non-linear function in the \textit{NAR} module.

\subsection{Baseline methods}

For this experiments, an extensive number of session-based recommendation algorithms was used for comparison.

Despite of the simplicity of some of those methods, they usually have competitive accuracy on session-based recommendations, for keeping up with the dynamic global reading interests and for being efficiently updated over time \cite{jannach2017recurrent} \cite{jugovac2018streamingrec} \cite{ludewig2018evaluation}.

\begin{itemize}
    \item \textbf{GRU4Rec} - Seminal neural architecture using RNNs for session-based recommendations \cite{hidasi2016}. For this experiment, it was used \textit{GRU4Rec} v2 implementation, with the improvements of \cite{hidasi2017} \footnote{The \textit{GRU4Rec} v2 \cite{hidasi2017} was released on Jun 12, 2017 and is available in https://github.com/hidasib/GRU4Rec}; 
    \item \textbf{Co-occurrent} - Recommends articles commonly viewed together with the last read article, in other user sessions. This algorithm is a simplified version of the association rules technique, with the maximum rule size of two (pairwise item co-occurrences) \cite{jugovac2018streamingrec} \cite{ludewig2018evaluation};
    \item \textbf{Sequential Rules (SR)} - A more sophisticated version of association rules, which considers the sequence of clicked items within the session. A rule is created when an item \textit{q} appeared after an item \textit{p} in a session, even when other items were viewed between \textit{p} and \textit{q}. The rules are weighted by the distance \textit{x} (number of steps) between \textit{p} and \textit{q} in the session with a linear weighting function $ w_{SR} = 1/x $ \cite{ludewig2018evaluation};
    \item \textbf{Item-kNN} - Returns most similar items to the last read article, in terms of the cosine similarity between the vector of their sessions, i.e. it is the number of co-occurrences of two items in sessions divided by the square root of the product of the numbers of sessions in which the individual items are occurred. This was the strongest baseline compared to \textit{GRU4Rec} in \cite{hidasi2016} \cite{ludewig2018evaluation};
    \item \textbf{Vector Multiplication Session-Based kNN (V-SkNN)} - Compares the entire active session with past sessions and find items to be recommended. The comparison emphasizes items more recently clicked within the session, when computing the similarities with past sessions \cite{jannach2017recurrent} \cite{jugovac2018streamingrec} \cite{ludewig2018evaluation};
    \item \textbf{Recently Popular} - Recommends the most viewed articles from the last N  clicks buffer; and
    \item \textbf{Content-Based} - For each article read by the user, recommends similar articles based on the cosine similarity of their \textit{Article Content Embeddings}, from the last N clicks buffer.
\end{itemize}

\subsubsection{Notes and improvements on GRU4Rec} 

The \textit{GRU4Rec} v2 \cite{hidasi2017} implementation have added the capability of incremental retraining, which supports training on sessions with new items ,without the need to retrain on the full dataset. In this setting, it dynamically adds input and output neuron units corresponding to new items, whose connections are randomly initialized.

Therefore, the \textit{GRU4Rec} does not support recommending items not seen on training. For this reason, during its evaluation, it was necessary to ignore fresh articles published since the last training (once each hour), which corresponded to about 2\% of ignored clicked items. This approach might have slightly overestimated \textit{GRU4Rec} accuracy, as recommending unknown items tends to be more challenging.

\textit{GRU4Rec} v2 \cite{hidasi2017} features an interesting negative sampling strategy, where it can be balanced between uniform sampling (all items have the same probability to be picked) or popularity-based sampling (item sampling probability is proportional to its support). 

For these experiments, \textit{GRU4Rec}'s uniform sampling lead to a better accuracy than popularity-based sampling. Therefore, as training evolved, it was possible to observe a decreasing accuracy, as the number of unique items increased, thus, the chance to randomly pick old articles as negative samples, which are not relevant anymore. 

For this reason, it was implemented for \textit{GRU4Rec} the same sampling strategy used in CHAMELEON during training, which uniformly samples negative items from the last N clicks buffer. With this improvement, it was possible to obtain a relative increase of 13\% on HR@5 and 18\% on MRR@5 for \textit{GRU4Rec}. 

The \textit{GRU4Rec} is very sensitive to hyperparameter choices, which were tuned in a separate period of the same dataset. The best architecture found for \textit{GRU4Rec} used a hidden layer with 300 units, BPR-max loss function, no item embeddings, and no dropout. It was trained with Adam optimizer (momentum=0), learning rate of 1e-4, $ L_2 $ regularization of 1e-5, and 200 negative samples (from the last N clicks buffer).

\subsection{Results}

Two experiments were performed in this study, involving different time periods and evaluation frequency:

\begin{enumerate}
    \item Continuous training and evaluating each five hours, during 15 days (Oct. 1-15, 2017); and
    \item Continuous training and evaluating each hour, on the subsequent day (Oct. 16, 2017).
\end{enumerate}

HR@5 and MRR@5 metrics were averaged and reported for each evaluation hour.

To keep results comparable, during the \textit{NAR} module evaluation, it was logged the sampled negative items for each session. Thus, the same negative samples by session were used for baseline methods evaluation.

\subsubsection{Experiment 1}
For this experiment, recommendation models were trained and evaluated on user sessions from a period of 15 days. Articles context attributes (recent popularity and recency) were also updated over time, to emulate a live environment.

After training on sessions corresponding to five hours, sessions within the subsequent hour were used for evaluation. 

\begin{figure}[h]
	\includegraphics[width=8cm]{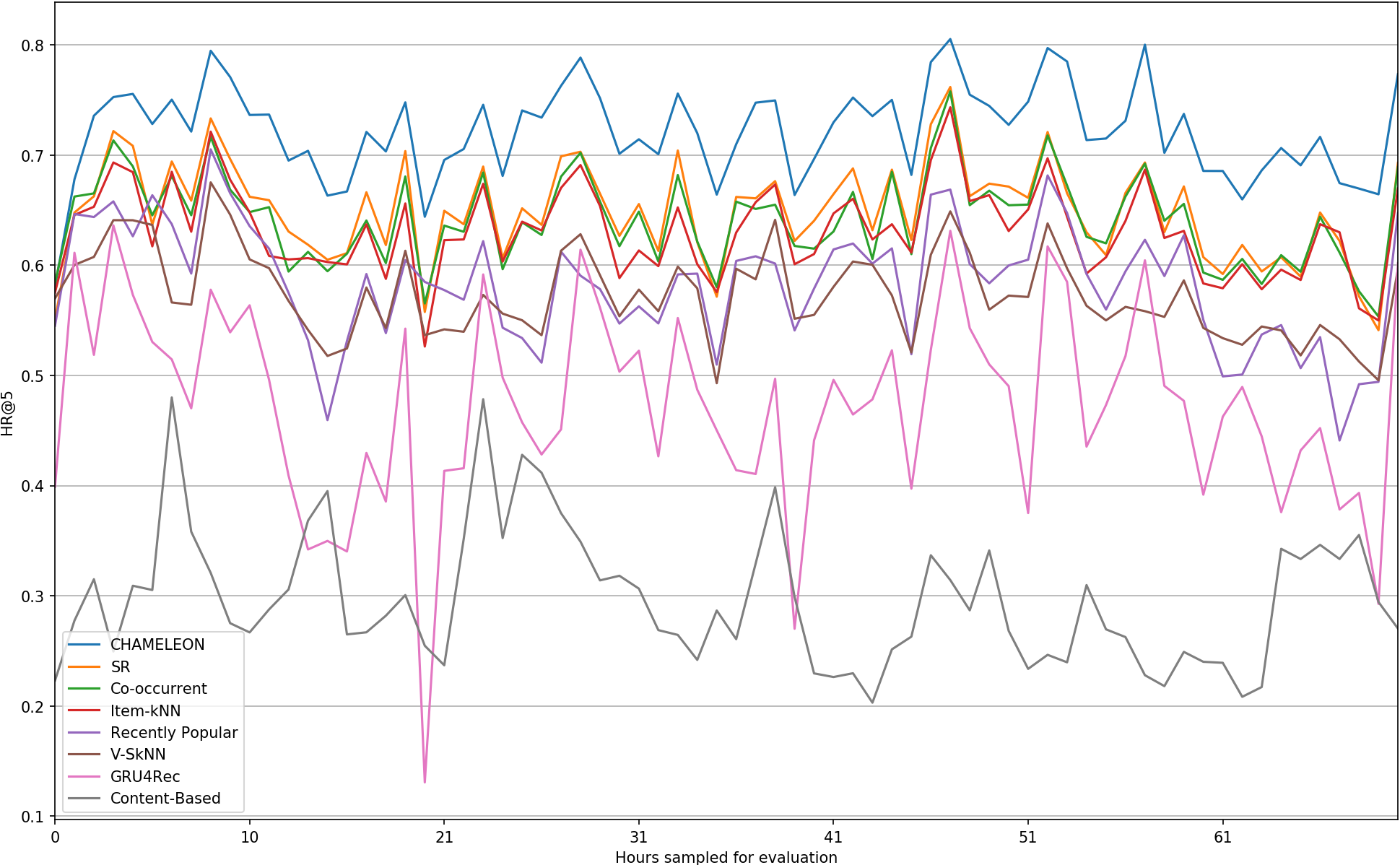}	\caption{Average HR@5 by hour (sampled for evaluation), for a 15-days period}
	\label{figure:hr_15_days_lines}
\end{figure}

\begin{figure}[h]
	\includegraphics[width=8cm]{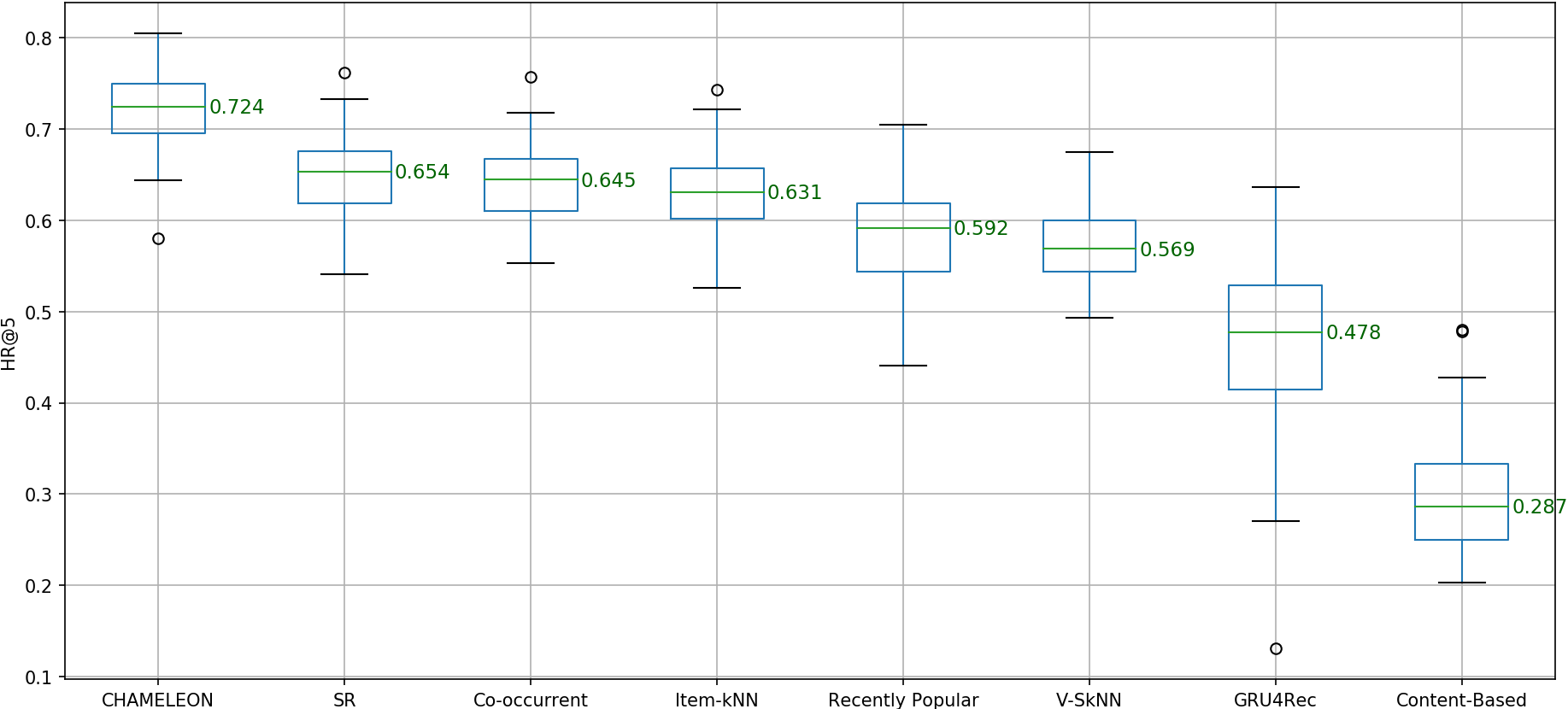}	\caption{Distribution of average HR@5 by hour (sampled for evaluation), for a 15-days period}
	\label{figure:hr_15_days_boxplot}
\end{figure}

Figure~\ref{figure:hr_15_days_lines} presents the evolution of HR@5 over time, for the sampled evaluation hours. The distribution of the average HR@5 by sampled hours can be seen in Figure~\ref{figure:hr_15_days_boxplot}. 

It can be observed in Figure~\ref{figure:hr_15_days_lines} that the proposed \textit{CHAMELEON} instantiation keeps constantly a higher accuracy than the baseline methods over time. As shown in Figure~\ref{figure:hr_15_days_boxplot}, the median HR@5 for \textit{CHAMELEON} was 0.72, while the best benchmark (SR) got a median of 0.65, a relative improvement of about 10\%.

Methods based on k-Nearest Neighbors and Association Rules presented a competitive accuracy on news session-based recommendations, even higher than \textit{GRU4Rec}, as also observed in \cite{jannach2017recurrent} \cite{jugovac2018streamingrec} \cite{ludewig2018evaluation}.

The lower accuracy was obtained by the Content-based method, which may indicate that textual similarity is not the best predictor for the next-read article.

\begin{figure}[h]
	\includegraphics[width=8cm]{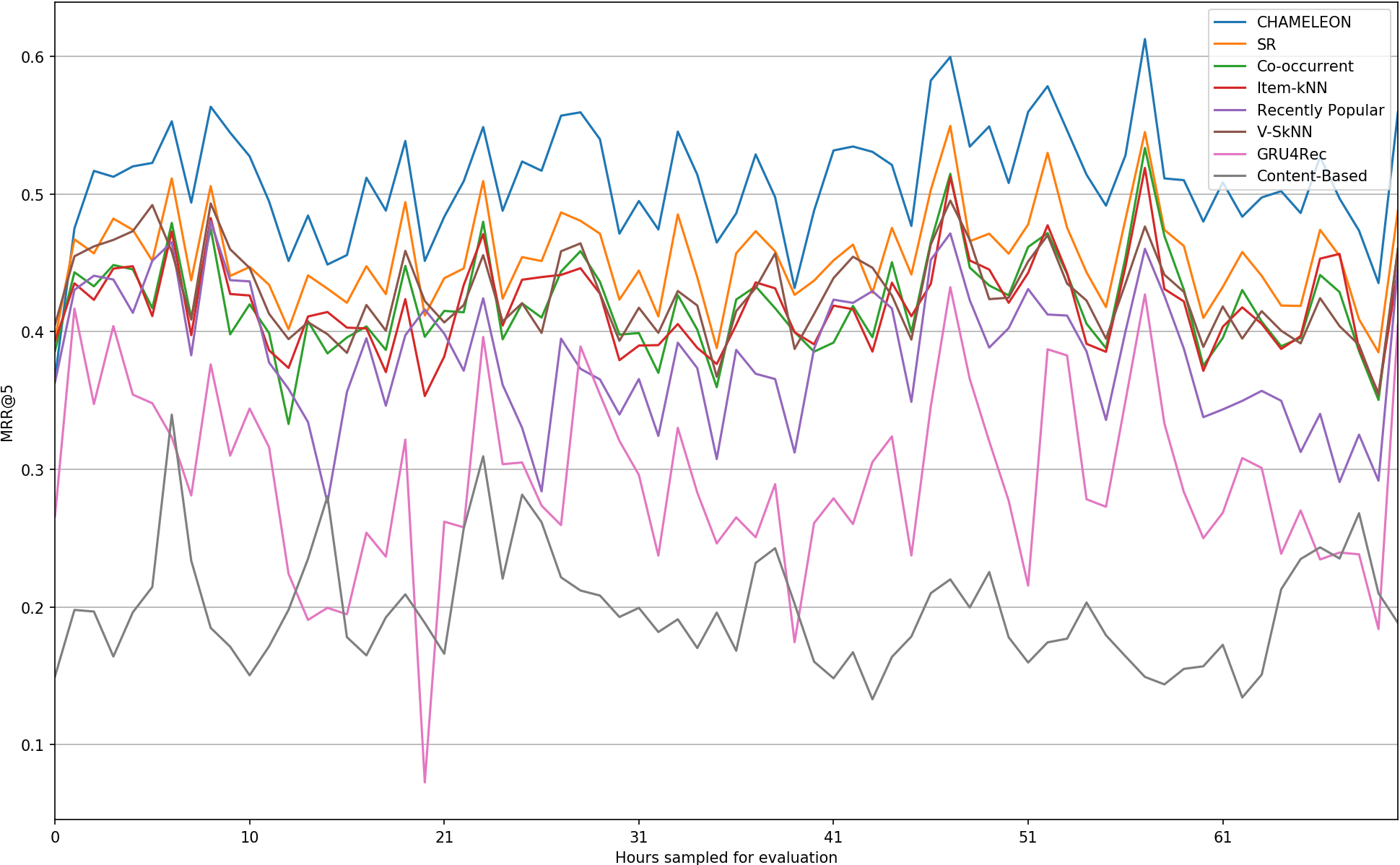}	\caption{Average MRR@5 by hour (sampled for evaluation), for a 15-days period}
	\label{figure:mrr_15_days_lines}
\end{figure}

\begin{figure}[h]
	\includegraphics[width=8cm]{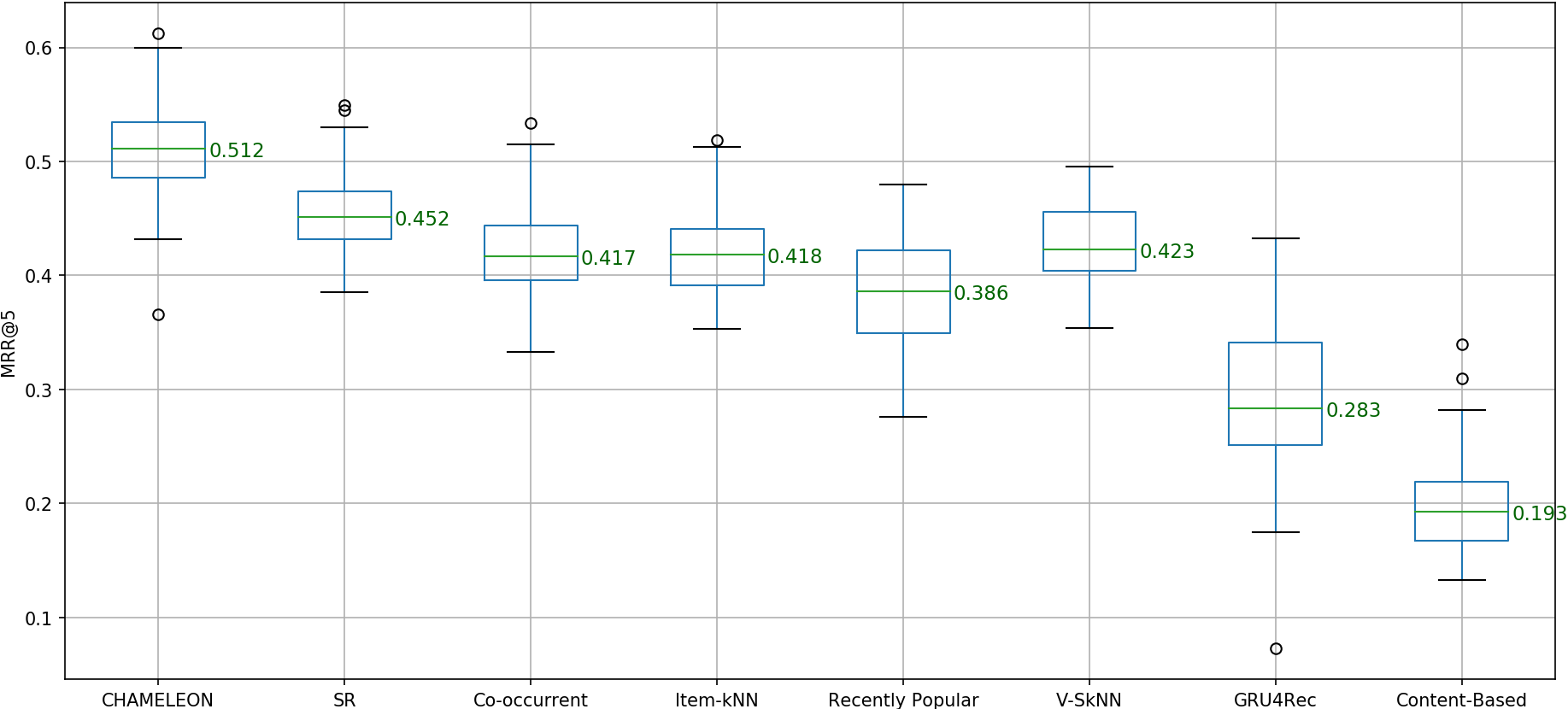}	\caption{Distribution of average MRR@5 by hour (sampled for evaluation), for a 15-days period}
	\label{figure:mrr_15_days_boxplot}
\end{figure}

The results are is similar for MRR@5. Figures~\ref{figure:mrr_15_days_lines}~and~\ref{figure:mrr_15_days_boxplot} show that the accuracy and ranking quality of recommendations provided by \textit{CHAMELEON} was constantly better than benchmarks methods. The median MRR@5 for \textit{CHAMELEON} was 0.51, while the best benchmark (SR) got a median of 0.45, a relative improvement of about 13\%.

\subsubsection{Experiment 2}
In this experiment, recommendation methods were evaluated more often (once an hour), for a period of 24 hours (Oct. 16, 2017). Thus, after training incrementally on sessions from each hour, sessions of the next hour were used for evaluation.

The models trained in \textit{Experiment 1} were used to initialize (parameters and states) models for \textit{Experiment 2}, to emulate a RS that has already being trained for some days.

\begin{figure}[h]
	\includegraphics[width=8cm]{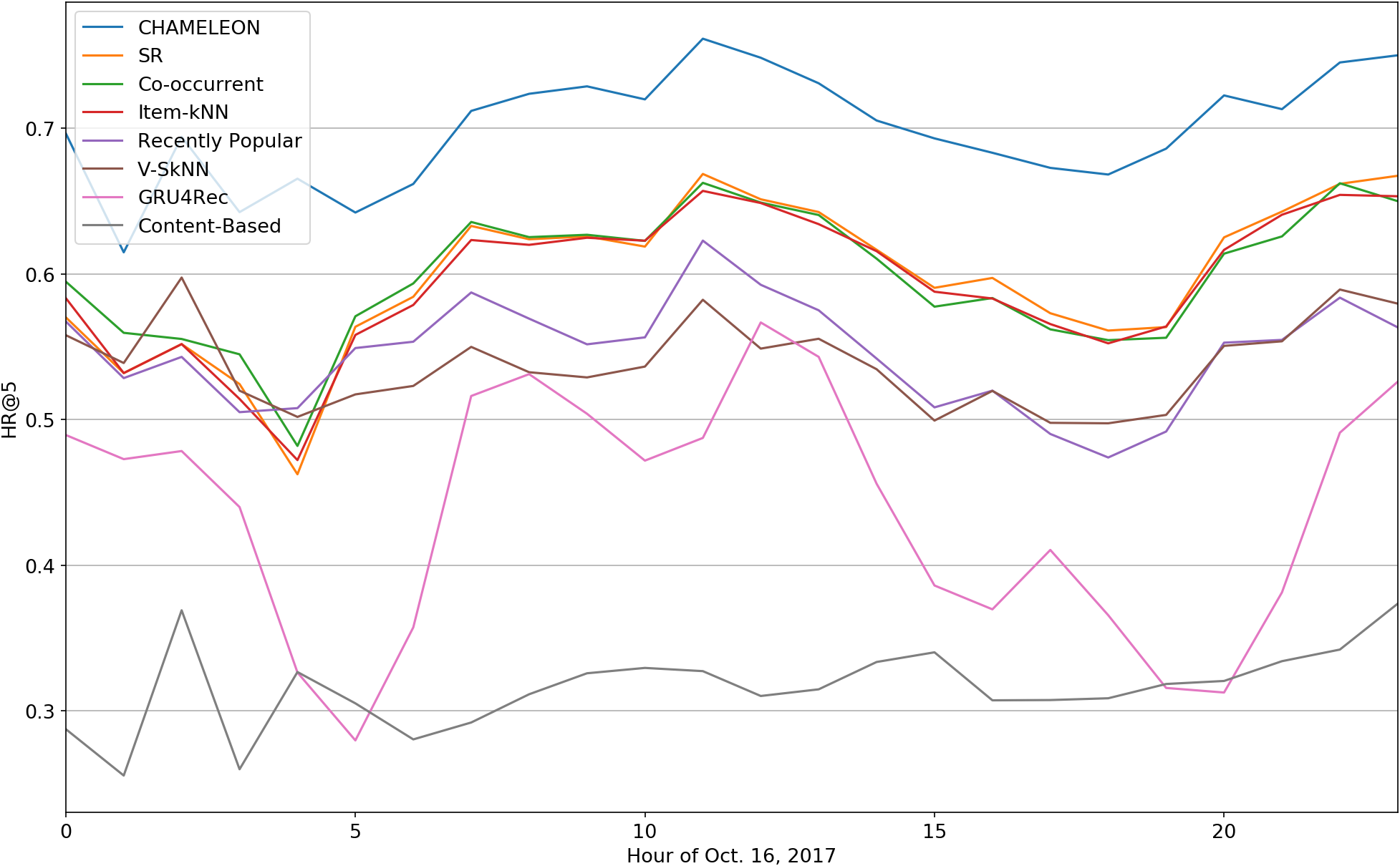}	\caption{Average HR@5 by hour, for Oct. 16, 2017}
	\label{figure:hitrate_at_5_last_day_lines}
\end{figure}

\begin{figure}[h]
	\includegraphics[width=8cm]{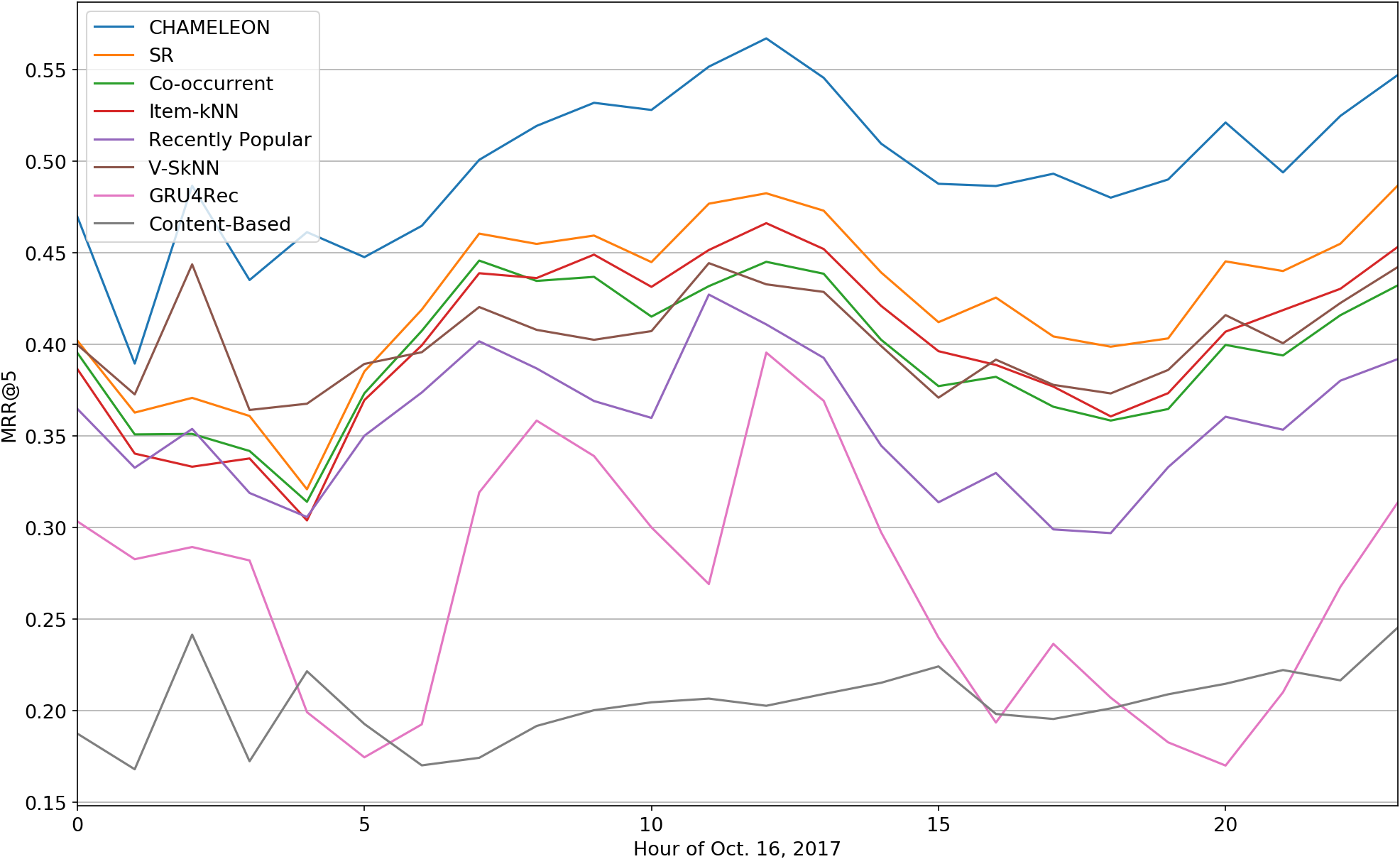}	\caption{Average MRR@5 by hour, for Oct. 16, 2017}
	\label{figure:mrr_at_5_last_day_lines}
\end{figure}

Figures~\ref{figure:hitrate_at_5_last_day_lines}~and~\ref{figure:mrr_at_5_last_day_lines} presents the evolution of the average HR@5 and MRR@5 by hour, within a period of 24 hours. Once again, it can be seen that the accuracy and ranking quality obtained by the proposed \textit{CHAMELEON} instantiation keeps higher than the baseline methods, throughout the day. 

\section{Related Work}

One of the main inspirations for the \textit{CHAMELEON} was the \textit{GRU4Rec} \cite{hidasi2016}, the seminal work on the usage of Recurrent Neural Networks (RNN) on session-based recommendations, and subsequent work \cite{hidasi2016parallel} \cite{hidasi2017}. 

Other main inspiration came from the \textit{Multi-View Deep Neural Network (MV-DNN)} \cite{elkahky2015multi}, which adapted \textit{Deep Structured Semantic Model (DSSM)} \cite{huang2013learning} for the recommendation task. 

The \textit{MV-DNN} maps users and items to a latent space, where the cosine similarity between users and their preferred items is maximized. That approach makes it possible to keep the neural network architecture static, rather than adding new units into the output layer for each new item (e.g., published article), as required by \textit{GRU4Rec} (softmax loss function) \cite{hidasi2016}. 

The \textit{MV-DNN} was adapted for news recommendation by \cite{song2016multi} \textit{Temporal DSSM (TDSSM)} and \cite{kumar2017} Recurrent Attention \textit{DSSM} (\textit{RA-DSSM}). 

Differently from \textit{CHAMELEON}, \textit{TDSSM} \cite{song2016multi} did not model user sessions explicitly, and items and users representations are not directly learned from news content and users behaviours. 

The \textit{RA-DSSM} \cite{kumar2017} represents articles content by using \textit{Doc2Vec} \cite{le2014distributed} embeddings (unsupervised training), while \textit{CHAMELEON} trains \textit{Article Content Embeddings} to predict news metadata by using supervised learning. The \textit{RA-DSSM} does not use any contextual information about the user and articles, which may limit its accuracy in a extreme cold-start scenario like news RS. Finally, the offline evaluation for that study has used cross-validation based on random sampling to predict the last article read by the user in a session, whilst for this study it was proposed a temporal offline evaluation method, to mimic a more realist scenario.

\section{Conclusion}
In this study, it was proposed an instantiation of the \textit{CHAMELEON} -- a Deep Learning Meta-Architecture for News Recommender Systems, using a CNN to extract textual features from news articles and a LSTM layer to model the sequence of clicked items in user sessions.

It was possible to observe that the recommendations accuracy obtained by the proposed \textit{CHAMELEON} instantiation was constantly higher over time than an extensive number of baseline methods for session-based recommendation, including the popular \textit{GRU4Rec}.

A temporal offline evaluation method was also proposed to emulate the dynamics of news readership, where articles context (recent popularity and recency) is constantly changing. Recommender methods are continuously trained on streaming user clicks and the model may be often deployed (e.g. once an hour) to a production environment, in order to serve recommendations for real users.

This is an ongoing research on news recommendation using the \textit{CHAMELEON} Deep Learning Meta-Architecture. Some of next planned steps are: (1) to understand how different input features (e.g., article content, user context, article context) individually impact the recommendations accuracy; (2) to evaluate recommendations of fresh news articles (item cold-start); (3) to explore other approaches to learn better article content embeddings; and (4) to evaluate models from perspectives other than accuracy, like popularity, coverage and serendipity, in different large news datasets.

It is suggested the adaptation of \textit{CHAMELEON} to provide contextual session-based recommendations in other domains, like e-commerce and media.

\begin{acks}
The authors would like to thank Globo.com for providing context on its challenges for large-scale news recommender systems and for sharing a dataset to make those experiments possible.
\end{acks}

\bibliographystyle{ACM-Reference-Format}
\bibliography{ms}


\begin{thebibliography}{70}


\ifx \showCODEN    \undefined \def \showCODEN     #1{\unskip}     \fi
\ifx \showDOI      \undefined \def \showDOI       #1{#1}\fi
\ifx \showISBNx    \undefined \def \showISBNx     #1{\unskip}     \fi
\ifx \showISBNxiii \undefined \def \showISBNxiii  #1{\unskip}     \fi
\ifx \showISSN     \undefined \def \showISSN      #1{\unskip}     \fi
\ifx \showLCCN     \undefined \def \showLCCN      #1{\unskip}     \fi
\ifx \shownote     \undefined \def \shownote      #1{#1}          \fi
\ifx \showarticletitle \undefined \def \showarticletitle #1{#1}   \fi
\ifx \showURL      \undefined \def \showURL       {\relax}        \fi
\providecommand\bibfield[2]{#2}
\providecommand\bibinfo[2]{#2}
\providecommand\natexlab[1]{#1}
\providecommand\showeprint[2][]{arXiv:#2}

\bibitem[\protect\citeauthoryear{Abadi, Agarwal, Barham, Brevdo, Chen, Citro,
  Corrado, Davis, Dean, Devin, et~al\mbox{.}}{Abadi et~al\mbox{.}}{2016}]%
        {abadi2016tensorflow}
\bibfield{author}{\bibinfo{person}{Mart{\'\i}n Abadi}, \bibinfo{person}{Ashish
  Agarwal}, \bibinfo{person}{Paul Barham}, \bibinfo{person}{Eugene Brevdo},
  \bibinfo{person}{Zhifeng Chen}, \bibinfo{person}{Craig Citro},
  \bibinfo{person}{Greg~S Corrado}, \bibinfo{person}{Andy Davis},
  \bibinfo{person}{Jeffrey Dean}, \bibinfo{person}{Matthieu Devin},
  {et~al\mbox{.}}} \bibinfo{year}{2016}\natexlab{}.
\newblock \showarticletitle{Tensorflow: Large-scale machine learning on
  heterogeneous distributed systems}.
\newblock \bibinfo{journal}{\emph{arXiv preprint arXiv:1603.04467}}
  (\bibinfo{year}{2016}).
\newblock


\bibitem[\protect\citeauthoryear{Ba, Kiros, and Hinton}{Ba
  et~al\mbox{.}}{2016}]%
        {ba2016layer}
\bibfield{author}{\bibinfo{person}{Jimmy~Lei Ba}, \bibinfo{person}{Jamie~Ryan
  Kiros}, {and} \bibinfo{person}{Geoffrey~E Hinton}.}
  \bibinfo{year}{2016}\natexlab{}.
\newblock \showarticletitle{Layer normalization}.
\newblock \bibinfo{journal}{\emph{arXiv preprint arXiv:1607.06450}}
  (\bibinfo{year}{2016}).
\newblock


\bibitem[\protect\citeauthoryear{Badenes, Bengualid, Chen, Gou, Haber, Mahmud,
  Nichols, Pal, Schoudt, Smith, et~al\mbox{.}}{Badenes et~al\mbox{.}}{2014}]%
        {Badenes2014}
\bibfield{author}{\bibinfo{person}{Hernan Badenes}, \bibinfo{person}{Mateo~N
  Bengualid}, \bibinfo{person}{Jilin Chen}, \bibinfo{person}{Liang Gou},
  \bibinfo{person}{Eben Haber}, \bibinfo{person}{Jalal Mahmud},
  \bibinfo{person}{Jeffrey~W Nichols}, \bibinfo{person}{Aditya Pal},
  \bibinfo{person}{Jerald Schoudt}, \bibinfo{person}{Barton~A Smith},
  {et~al\mbox{.}}} \bibinfo{year}{2014}\natexlab{}.
\newblock \showarticletitle{System U: automatically deriving personality traits
  from social media for people recommendation}. In
  \bibinfo{booktitle}{\emph{Proceedings of the 8th ACM Conference on
  Recommender systems}}. ACM, \bibinfo{pages}{373--374}.
\newblock


\bibitem[\protect\citeauthoryear{Bansal, Belanger, and McCallum}{Bansal
  et~al\mbox{.}}{2016}]%
        {bansal2016ask}
\bibfield{author}{\bibinfo{person}{Trapit Bansal}, \bibinfo{person}{David
  Belanger}, {and} \bibinfo{person}{Andrew McCallum}.}
  \bibinfo{year}{2016}\natexlab{}.
\newblock \showarticletitle{Ask the gru: Multi-task learning for deep text
  recommendations}. In \bibinfo{booktitle}{\emph{Proceedings of the 10th ACM
  Conference on Recommender Systems}}. ACM, \bibinfo{pages}{107--114}.
\newblock


\bibitem[\protect\citeauthoryear{Bastian, Hayes, Vaughan, Shah, Skomoroch, Kim,
  Uryasev, and Lloyd}{Bastian et~al\mbox{.}}{2014}]%
        {Bastian2014}
\bibfield{author}{\bibinfo{person}{Mathieu Bastian}, \bibinfo{person}{Matthew
  Hayes}, \bibinfo{person}{William Vaughan}, \bibinfo{person}{Sam Shah},
  \bibinfo{person}{Peter Skomoroch}, \bibinfo{person}{Hyungjin Kim},
  \bibinfo{person}{Sal Uryasev}, {and} \bibinfo{person}{Christopher Lloyd}.}
  \bibinfo{year}{2014}\natexlab{}.
\newblock \showarticletitle{LinkedIn skills: large-scale topic extraction and
  inference}. In \bibinfo{booktitle}{\emph{Proceedings of the 8th ACM
  Conference on Recommender systems}}. ACM, \bibinfo{pages}{1--8}.
\newblock


\bibitem[\protect\citeauthoryear{Beel, Langer, Genzmehr, Gipp, Breitinger, and
  N{\"u}rnberger}{Beel et~al\mbox{.}}{2013b}]%
        {Beel2013b}
\bibfield{author}{\bibinfo{person}{Joeran Beel}, \bibinfo{person}{Stefan
  Langer}, \bibinfo{person}{Marcel Genzmehr}, \bibinfo{person}{Bela Gipp},
  \bibinfo{person}{Corinna Breitinger}, {and} \bibinfo{person}{Andreas
  N{\"u}rnberger}.} \bibinfo{year}{2013}\natexlab{b}.
\newblock \showarticletitle{Research paper recommender system evaluation: A
  quantitative literature survey}. In \bibinfo{booktitle}{\emph{Proceedings of
  the International Workshop on Reproducibility and Replication in Recommender
  Systems Evaluation}}. ACM, \bibinfo{pages}{15--22}.
\newblock


\bibitem[\protect\citeauthoryear{Beel, Langer, Genzmehr, and
  N{\"u}rnberger}{Beel et~al\mbox{.}}{2013a}]%
        {beel2013introducing}
\bibfield{author}{\bibinfo{person}{Joeran Beel}, \bibinfo{person}{Stefan
  Langer}, \bibinfo{person}{Marcel Genzmehr}, {and} \bibinfo{person}{Andreas
  N{\"u}rnberger}.} \bibinfo{year}{2013}\natexlab{a}.
\newblock \showarticletitle{Introducing Docear's research paper recommender
  system}. In \bibinfo{booktitle}{\emph{Proceedings of the 13th ACM/IEEE-CS
  joint conference on Digital libraries}}. ACM, \bibinfo{pages}{459--460}.
\newblock


\bibitem[\protect\citeauthoryear{Bengio et~al\mbox{.}}{Bengio
  et~al\mbox{.}}{2009}]%
        {bengio2009}
\bibfield{author}{\bibinfo{person}{Yoshua Bengio} {et~al\mbox{.}}}
  \bibinfo{year}{2009}\natexlab{}.
\newblock \showarticletitle{Learning deep architectures for AI}.
\newblock \bibinfo{journal}{\emph{Foundations and trends{\textregistered} in
  Machine Learning}} \bibinfo{volume}{2}, \bibinfo{number}{1}
  (\bibinfo{year}{2009}), \bibinfo{pages}{1--127}.
\newblock


\bibitem[\protect\citeauthoryear{Bengio, LeCun, et~al\mbox{.}}{Bengio
  et~al\mbox{.}}{2007}]%
        {bengio2007}
\bibfield{author}{\bibinfo{person}{Yoshua Bengio}, \bibinfo{person}{Yann
  LeCun}, {et~al\mbox{.}}} \bibinfo{year}{2007}\natexlab{}.
\newblock \showarticletitle{Scaling learning algorithms towards AI}.
\newblock \bibinfo{journal}{\emph{Large-scale kernel machines}}
  \bibinfo{volume}{34}, \bibinfo{number}{5} (\bibinfo{year}{2007}),
  \bibinfo{pages}{1--41}.
\newblock


\bibitem[\protect\citeauthoryear{Bilton}{Bilton}{2016}]%
        {bilton2016}
\bibfield{author}{\bibinfo{person}{R. Bilton}.}
  \bibinfo{year}{2016}\natexlab{}.
\newblock \bibinfo{title}{The Washington Post tests personalized "pop-up"
  newsletters to promote its big stories}.
\newblock
  \bibinfo{howpublished}{\url{http://www.niemanlab.org/2016/05/the-washington-post-tests-personalized-pop-up-newsletters-to-promote-its-big-stories/}}.
    (\bibinfo{date}{May} \bibinfo{year}{2016}).
\newblock


\bibitem[\protect\citeauthoryear{Bu, Tan, Chen, Wang, Wu, Zhang, and He}{Bu
  et~al\mbox{.}}{2010}]%
        {Bu2010}
\bibfield{author}{\bibinfo{person}{Jiajun Bu}, \bibinfo{person}{Shulong Tan},
  \bibinfo{person}{Chun Chen}, \bibinfo{person}{Can Wang}, \bibinfo{person}{Hao
  Wu}, \bibinfo{person}{Lijun Zhang}, {and} \bibinfo{person}{Xiaofei He}.}
  \bibinfo{year}{2010}\natexlab{}.
\newblock \showarticletitle{Music recommendation by unified hypergraph:
  combining social media information and music content}. In
  \bibinfo{booktitle}{\emph{Proceedings of the international conference on
  Multimedia}}. ACM, \bibinfo{pages}{391--400}.
\newblock


\bibitem[\protect\citeauthoryear{Capelle, Frasincar, Moerland, and
  Hogenboom}{Capelle et~al\mbox{.}}{2012}]%
        {capelle2012semantics}
\bibfield{author}{\bibinfo{person}{Michel Capelle}, \bibinfo{person}{Flavius
  Frasincar}, \bibinfo{person}{Marnix Moerland}, {and}
  \bibinfo{person}{Frederik Hogenboom}.} \bibinfo{year}{2012}\natexlab{}.
\newblock \showarticletitle{Semantics-based news recommendation}. In
  \bibinfo{booktitle}{\emph{Proceedings of the 2nd international conference on
  web intelligence, mining and semantics}}. ACM, \bibinfo{pages}{27}.
\newblock


\bibitem[\protect\citeauthoryear{Catherine and Cohen}{Catherine and
  Cohen}{2017}]%
        {catherine2017transnets}
\bibfield{author}{\bibinfo{person}{Rose Catherine} {and}
  \bibinfo{person}{William Cohen}.} \bibinfo{year}{2017}\natexlab{}.
\newblock \showarticletitle{TransNets: Learning to Transform for
  Recommendation}.
\newblock \bibinfo{journal}{\emph{arXiv preprint arXiv:1704.02298}}
  (\bibinfo{year}{2017}).
\newblock


\bibitem[\protect\citeauthoryear{Chu and Park}{Chu and Park}{2009}]%
        {chu2009personalized}
\bibfield{author}{\bibinfo{person}{Wei Chu} {and} \bibinfo{person}{Seung-Taek
  Park}.} \bibinfo{year}{2009}\natexlab{}.
\newblock \showarticletitle{Personalized recommendation on dynamic content
  using predictive bilinear models}. In \bibinfo{booktitle}{\emph{Proceedings
  of the 18th international conference on World wide web}}. ACM,
  \bibinfo{pages}{691--700}.
\newblock


\bibitem[\protect\citeauthoryear{Das, Datar, Garg, and Rajaram}{Das
  et~al\mbox{.}}{2007}]%
        {das2007}
\bibfield{author}{\bibinfo{person}{Abhinandan~S Das}, \bibinfo{person}{Mayur
  Datar}, \bibinfo{person}{Ashutosh Garg}, {and} \bibinfo{person}{Shyam
  Rajaram}.} \bibinfo{year}{2007}\natexlab{}.
\newblock \showarticletitle{Google news personalization: scalable online
  collaborative filtering}. In \bibinfo{booktitle}{\emph{Proceedings of the
  16th international conference on World Wide Web}}. ACM,
  \bibinfo{pages}{271--280}.
\newblock


\bibitem[\protect\citeauthoryear{Davidson, Liebald, Liu, Nandy, Van~Vleet,
  Gargi, Gupta, He, Lambert, Livingston, et~al\mbox{.}}{Davidson
  et~al\mbox{.}}{2010}]%
        {davidson2010youtube}
\bibfield{author}{\bibinfo{person}{James Davidson}, \bibinfo{person}{Benjamin
  Liebald}, \bibinfo{person}{Junning Liu}, \bibinfo{person}{Palash Nandy},
  \bibinfo{person}{Taylor Van~Vleet}, \bibinfo{person}{Ullas Gargi},
  \bibinfo{person}{Sujoy Gupta}, \bibinfo{person}{Yu He}, \bibinfo{person}{Mike
  Lambert}, \bibinfo{person}{Blake Livingston}, {et~al\mbox{.}}}
  \bibinfo{year}{2010}\natexlab{}.
\newblock \showarticletitle{The YouTube video recommendation system}. In
  \bibinfo{booktitle}{\emph{Proceedings of the fourth ACM conference on
  Recommender systems}}. ACM, \bibinfo{pages}{293--296}.
\newblock


\bibitem[\protect\citeauthoryear{D{\'\i}ez~Pel{\'a}ez, Mart{\'\i}nez~Rego,
  Alonso~Betanzos, Luaces~Rodr{\'\i}guez, and
  Bahamonde~Rionda}{D{\'\i}ez~Pel{\'a}ez et~al\mbox{.}}{2016}]%
        {diez2016}
\bibfield{author}{\bibinfo{person}{Jorge D{\'\i}ez~Pel{\'a}ez},
  \bibinfo{person}{David Mart{\'\i}nez~Rego}, \bibinfo{person}{Amparo
  Alonso~Betanzos}, \bibinfo{person}{{\'O}scar Luaces~Rodr{\'\i}guez}, {and}
  \bibinfo{person}{Antonio Bahamonde~Rionda}.} \bibinfo{year}{2016}\natexlab{}.
\newblock \showarticletitle{Metrical Representation of Readers and Articles in
  a Digital Newspaper}. In \bibinfo{booktitle}{\emph{10th ACM Conference on
  Recommender Systems (RecSys 2016)}}. ACM.
\newblock


\bibitem[\protect\citeauthoryear{Donkers, Loepp, and Ziegler}{Donkers
  et~al\mbox{.}}{2017}]%
        {donkers2017sequential}
\bibfield{author}{\bibinfo{person}{Tim Donkers}, \bibinfo{person}{Benedikt
  Loepp}, {and} \bibinfo{person}{J{\"u}rgen Ziegler}.}
  \bibinfo{year}{2017}\natexlab{}.
\newblock \showarticletitle{Sequential User-based Recurrent Neural Network
  Recommendations}. In \bibinfo{booktitle}{\emph{Proceedings of the Eleventh
  ACM Conference on Recommender Systems}}. ACM, \bibinfo{pages}{152--160}.
\newblock


\bibitem[\protect\citeauthoryear{Elkahky, Song, and He}{Elkahky
  et~al\mbox{.}}{2015}]%
        {elkahky2015multi}
\bibfield{author}{\bibinfo{person}{Ali~Mamdouh Elkahky}, \bibinfo{person}{Yang
  Song}, {and} \bibinfo{person}{Xiaodong He}.} \bibinfo{year}{2015}\natexlab{}.
\newblock \showarticletitle{A multi-view deep learning approach for cross
  domain user modeling in recommendation systems}. In
  \bibinfo{booktitle}{\emph{Proceedings of the 24th International Conference on
  World Wide Web}}. International World Wide Web Conferences Steering
  Committee, \bibinfo{pages}{278--288}.
\newblock


\bibitem[\protect\citeauthoryear{Epure, Kille, Ingvaldsen, Deneckere, Salinesi,
  and Albayrak}{Epure et~al\mbox{.}}{2017}]%
        {epure2017recommending}
\bibfield{author}{\bibinfo{person}{Elena~Viorica Epure},
  \bibinfo{person}{Benjamin Kille}, \bibinfo{person}{Jon~Espen Ingvaldsen},
  \bibinfo{person}{Rebecca Deneckere}, \bibinfo{person}{Camille Salinesi},
  {and} \bibinfo{person}{Sahin Albayrak}.} \bibinfo{year}{2017}\natexlab{}.
\newblock \showarticletitle{Recommending Personalized News in Short User
  Sessions}. In \bibinfo{booktitle}{\emph{Proceedings of the Eleventh ACM
  Conference on Recommender Systems}}. ACM, \bibinfo{pages}{121--129}.
\newblock


\bibitem[\protect\citeauthoryear{Glorot and Bengio}{Glorot and Bengio}{2010}]%
        {glorot2010understanding}
\bibfield{author}{\bibinfo{person}{Xavier Glorot} {and} \bibinfo{person}{Yoshua
  Bengio}.} \bibinfo{year}{2010}\natexlab{}.
\newblock \showarticletitle{Understanding the difficulty of training deep
  feedforward neural networks}. In \bibinfo{booktitle}{\emph{Proceedings of the
  thirteenth international conference on artificial intelligence and
  statistics}}. \bibinfo{pages}{249--256}.
\newblock


\bibitem[\protect\citeauthoryear{Gulla, Marco, Fidjest{\o}l, Ingvaldsen, and
  {\"O}zg{\"o}bek}{Gulla et~al\mbox{.}}{2016}]%
        {gulla2016intricacies}
\bibfield{author}{\bibinfo{person}{Jon~Atle Gulla}, \bibinfo{person}{Cristina
  Marco}, \bibinfo{person}{Arne~Dag Fidjest{\o}l}, \bibinfo{person}{Jon~Espen
  Ingvaldsen}, {and} \bibinfo{person}{{\"O}zlem {\"O}zg{\"o}bek}.}
  \bibinfo{year}{2016}\natexlab{}.
\newblock \showarticletitle{The Intricacies of Time in News Recommendation.}.
  In \bibinfo{booktitle}{\emph{UMAP (Extended Proceedings)}}.
\newblock


\bibitem[\protect\citeauthoryear{He, Zhang, Ren, and Sun}{He
  et~al\mbox{.}}{2016}]%
        {he2016deep}
\bibfield{author}{\bibinfo{person}{Kaiming He}, \bibinfo{person}{Xiangyu
  Zhang}, \bibinfo{person}{Shaoqing Ren}, {and} \bibinfo{person}{Jian Sun}.}
  \bibinfo{year}{2016}\natexlab{}.
\newblock \showarticletitle{Deep residual learning for image recognition}. In
  \bibinfo{booktitle}{\emph{Proceedings of the IEEE conference on computer
  vision and pattern recognition}}. \bibinfo{pages}{770--778}.
\newblock


\bibitem[\protect\citeauthoryear{Hidasi and Karatzoglou}{Hidasi and
  Karatzoglou}{2017}]%
        {hidasi2017}
\bibfield{author}{\bibinfo{person}{Bal{\'a}zs Hidasi} {and}
  \bibinfo{person}{Alexandros Karatzoglou}.} \bibinfo{year}{2017}\natexlab{}.
\newblock \showarticletitle{Recurrent Neural Networks with Top-k Gains for
  Session-based Recommendations}.
\newblock \bibinfo{journal}{\emph{arXiv preprint arXiv:1706.03847}}
  (\bibinfo{year}{2017}).
\newblock


\bibitem[\protect\citeauthoryear{Hidasi, Karatzoglou, Baltrunas, and
  Tikk}{Hidasi et~al\mbox{.}}{2016a}]%
        {hidasi2016}
\bibfield{author}{\bibinfo{person}{Bal{\'a}zs Hidasi},
  \bibinfo{person}{Alexandros Karatzoglou}, \bibinfo{person}{Linas Baltrunas},
  {and} \bibinfo{person}{Domonkos Tikk}.} \bibinfo{year}{2016}\natexlab{a}.
\newblock \showarticletitle{Session-based recommendations with recurrent neural
  networks}. In \bibinfo{booktitle}{\emph{Proceedings of Forth International
  Conference on Learning Representations}}.
\newblock


\bibitem[\protect\citeauthoryear{Hidasi, Karatzoglou, Sar-Shalom, Dieleman,
  Shapira, and Tikk}{Hidasi et~al\mbox{.}}{2017}]%
        {hidasi2017dlrs}
\bibfield{author}{\bibinfo{person}{Bal{\'a}zs Hidasi},
  \bibinfo{person}{Alexandros Karatzoglou}, \bibinfo{person}{Oren Sar-Shalom},
  \bibinfo{person}{Sander Dieleman}, \bibinfo{person}{Bracha Shapira}, {and}
  \bibinfo{person}{Domonkos Tikk}.} \bibinfo{year}{2017}\natexlab{}.
\newblock \showarticletitle{DLRS 2017-Second Workshop on Deep Learning for
  Recommender Systems}. In \bibinfo{booktitle}{\emph{Proceedings of the 2st
  Workshop on Deep Learning for Recommender Systems. ACM}},
  Vol.~\bibinfo{volume}{29}. \bibinfo{pages}{34}.
\newblock


\bibitem[\protect\citeauthoryear{Hidasi, Quadrana, Karatzoglou, and
  Tikk}{Hidasi et~al\mbox{.}}{2016b}]%
        {hidasi2016parallel}
\bibfield{author}{\bibinfo{person}{Bal{\'a}zs Hidasi}, \bibinfo{person}{Massimo
  Quadrana}, \bibinfo{person}{Alexandros Karatzoglou}, {and}
  \bibinfo{person}{Domonkos Tikk}.} \bibinfo{year}{2016}\natexlab{b}.
\newblock \showarticletitle{Parallel recurrent neural network architectures for
  feature-rich session-based recommendations}. In
  \bibinfo{booktitle}{\emph{Proceedings of the 10th ACM Conference on
  Recommender Systems}}. ACM, \bibinfo{pages}{241--248}.
\newblock


\bibitem[\protect\citeauthoryear{Hinton, Osindero, and Teh}{Hinton
  et~al\mbox{.}}{2006}]%
        {hinton2006}
\bibfield{author}{\bibinfo{person}{Geoffrey~E Hinton}, \bibinfo{person}{Simon
  Osindero}, {and} \bibinfo{person}{Yee-Whye Teh}.}
  \bibinfo{year}{2006}\natexlab{}.
\newblock \showarticletitle{A fast learning algorithm for deep belief nets}.
\newblock \bibinfo{journal}{\emph{Neural computation}} \bibinfo{volume}{18},
  \bibinfo{number}{7} (\bibinfo{year}{2006}), \bibinfo{pages}{1527--1554}.
\newblock


\bibitem[\protect\citeauthoryear{Hinton and Salakhutdinov}{Hinton and
  Salakhutdinov}{2006}]%
        {hinton2006b}
\bibfield{author}{\bibinfo{person}{Geoffrey~E Hinton} {and}
  \bibinfo{person}{Ruslan~R Salakhutdinov}.} \bibinfo{year}{2006}\natexlab{}.
\newblock \showarticletitle{Reducing the dimensionality of data with neural
  networks}.
\newblock \bibinfo{journal}{\emph{science}} \bibinfo{volume}{313},
  \bibinfo{number}{5786} (\bibinfo{year}{2006}), \bibinfo{pages}{504--507}.
\newblock


\bibitem[\protect\citeauthoryear{Hochreiter and Schmidhuber}{Hochreiter and
  Schmidhuber}{1997}]%
        {hochreiter1997long}
\bibfield{author}{\bibinfo{person}{Sepp Hochreiter} {and}
  \bibinfo{person}{J{\"u}rgen Schmidhuber}.} \bibinfo{year}{1997}\natexlab{}.
\newblock \showarticletitle{Long short-term memory}.
\newblock \bibinfo{journal}{\emph{Neural computation}} \bibinfo{volume}{9},
  \bibinfo{number}{8} (\bibinfo{year}{1997}), \bibinfo{pages}{1735--1780}.
\newblock


\bibitem[\protect\citeauthoryear{Huang, He, Gao, Deng, Acero, and Heck}{Huang
  et~al\mbox{.}}{2013}]%
        {huang2013learning}
\bibfield{author}{\bibinfo{person}{Po-Sen Huang}, \bibinfo{person}{Xiaodong
  He}, \bibinfo{person}{Jianfeng Gao}, \bibinfo{person}{Li Deng},
  \bibinfo{person}{Alex Acero}, {and} \bibinfo{person}{Larry Heck}.}
  \bibinfo{year}{2013}\natexlab{}.
\newblock \showarticletitle{Learning deep structured semantic models for web
  search using clickthrough data}. In \bibinfo{booktitle}{\emph{Proceedings of
  the 22nd ACM international conference on Conference on information \&
  knowledge management}}. ACM, \bibinfo{pages}{2333--2338}.
\newblock


\bibitem[\protect\citeauthoryear{Ilievski and Roy}{Ilievski and Roy}{2013}]%
        {ilievski2013personalized}
\bibfield{author}{\bibinfo{person}{Ilija Ilievski} {and} \bibinfo{person}{Sujoy
  Roy}.} \bibinfo{year}{2013}\natexlab{}.
\newblock \showarticletitle{Personalized news recommendation based on implicit
  feedback}. In \bibinfo{booktitle}{\emph{Proceedings of the 2013 international
  news recommender systems workshop and challenge}}. ACM,
  \bibinfo{pages}{10--15}.
\newblock


\bibitem[\protect\citeauthoryear{Jannach and Ludewig}{Jannach and
  Ludewig}{2017}]%
        {jannach2017recurrent}
\bibfield{author}{\bibinfo{person}{Dietmar Jannach} {and}
  \bibinfo{person}{Malte Ludewig}.} \bibinfo{year}{2017}\natexlab{}.
\newblock \showarticletitle{When recurrent neural networks meet the
  neighborhood for session-based recommendation}. In
  \bibinfo{booktitle}{\emph{Proceedings of the Eleventh ACM Conference on
  Recommender Systems}}. ACM, \bibinfo{pages}{306--310}.
\newblock


\bibitem[\protect\citeauthoryear{Jawaheer, Weller, and Kostkova}{Jawaheer
  et~al\mbox{.}}{2014}]%
        {jawaheer2014modeling}
\bibfield{author}{\bibinfo{person}{Gawesh Jawaheer}, \bibinfo{person}{Peter
  Weller}, {and} \bibinfo{person}{Patty Kostkova}.}
  \bibinfo{year}{2014}\natexlab{}.
\newblock \showarticletitle{Modeling user preferences in recommender systems: A
  classification framework for explicit and implicit user feedback}.
\newblock \bibinfo{journal}{\emph{ACM Transactions on Interactive Intelligent
  Systems (TiiS)}} \bibinfo{volume}{4}, \bibinfo{number}{2}
  (\bibinfo{year}{2014}), \bibinfo{pages}{8}.
\newblock


\bibitem[\protect\citeauthoryear{Jugovac, Jannach, and Karimi}{Jugovac
  et~al\mbox{.}}{2018}]%
        {jugovac2018streamingrec}
\bibfield{author}{\bibinfo{person}{Michael Jugovac}, \bibinfo{person}{Dietmar
  Jannach}, {and} \bibinfo{person}{Mozhgan Karimi}.}
  \bibinfo{year}{2018}\natexlab{}.
\newblock \showarticletitle{StreamingRec: A Framework for Benchmarking
  Stream-based News Recommenders}.
\newblock  (\bibinfo{year}{2018}), \bibinfo{pages}{306--310}.
\newblock


\bibitem[\protect\citeauthoryear{Kenthapadi, Le, and Venkataraman}{Kenthapadi
  et~al\mbox{.}}{2017}]%
        {kenthapadi2017personalized}
\bibfield{author}{\bibinfo{person}{Krishnaram Kenthapadi},
  \bibinfo{person}{Benjamin Le}, {and} \bibinfo{person}{Ganesh Venkataraman}.}
  \bibinfo{year}{2017}\natexlab{}.
\newblock \showarticletitle{Personalized Job Recommendation System at LinkedIn:
  Practical Challenges and Lessons Learned}. In
  \bibinfo{booktitle}{\emph{Proceedings of the Eleventh ACM Conference on
  Recommender Systems}}. ACM, \bibinfo{pages}{346--347}.
\newblock


\bibitem[\protect\citeauthoryear{Kingma and Ba}{Kingma and Ba}{2014}]%
        {kingma2014adam}
\bibfield{author}{\bibinfo{person}{Diederik~P Kingma} {and}
  \bibinfo{person}{Jimmy Ba}.} \bibinfo{year}{2014}\natexlab{}.
\newblock \showarticletitle{Adam: A method for stochastic optimization}.
\newblock \bibinfo{journal}{\emph{arXiv preprint arXiv:1412.6980}}
  (\bibinfo{year}{2014}).
\newblock


\bibitem[\protect\citeauthoryear{Kumar, Khattar, Gupta, Gupta, and Varma}{Kumar
  et~al\mbox{.}}{2017}]%
        {kumar2017}
\bibfield{author}{\bibinfo{person}{Vaibhav Kumar}, \bibinfo{person}{Dhruv
  Khattar}, \bibinfo{person}{Shashank Gupta}, \bibinfo{person}{Manish Gupta},
  {and} \bibinfo{person}{Vasudeva Varma}.} \bibinfo{year}{2017}\natexlab{}.
\newblock \showarticletitle{Deep Neural Architecture for News Recommendation}.
  In \bibinfo{booktitle}{\emph{Working Notes of the 8th International
  Conference of the CLEF Initiative, Dublin, Ireland. CEUR Workshop
  Proceedings}}.
\newblock


\bibitem[\protect\citeauthoryear{Le and Mikolov}{Le and Mikolov}{2014}]%
        {le2014distributed}
\bibfield{author}{\bibinfo{person}{Quoc Le} {and} \bibinfo{person}{Tomas
  Mikolov}.} \bibinfo{year}{2014}\natexlab{}.
\newblock \showarticletitle{Distributed representations of sentences and
  documents}. In \bibinfo{booktitle}{\emph{Proceedings of the 31st
  International Conference on Machine Learning (ICML-14)}}.
  \bibinfo{pages}{1188--1196}.
\newblock


\bibitem[\protect\citeauthoryear{Lee, Ahn, Lee, Ha, and Lee}{Lee
  et~al\mbox{.}}{2016}]%
        {lee2016quote}
\bibfield{author}{\bibinfo{person}{Hanbit Lee}, \bibinfo{person}{Yeonchan Ahn},
  \bibinfo{person}{Haejun Lee}, \bibinfo{person}{Seungdo Ha}, {and}
  \bibinfo{person}{Sang-goo Lee}.} \bibinfo{year}{2016}\natexlab{}.
\newblock \showarticletitle{Quote Recommendation in Dialogue using Deep Neural
  Network}. In \bibinfo{booktitle}{\emph{Proceedings of the 39th International
  ACM SIGIR conference on Research and Development in Information Retrieval}}.
  ACM, \bibinfo{pages}{957--960}.
\newblock


\bibitem[\protect\citeauthoryear{Li, Wang, Li, Knox, and Padmanabhan}{Li
  et~al\mbox{.}}{2011}]%
        {li2011scene}
\bibfield{author}{\bibinfo{person}{Lei Li}, \bibinfo{person}{Dingding Wang},
  \bibinfo{person}{Tao Li}, \bibinfo{person}{Daniel Knox}, {and}
  \bibinfo{person}{Balaji Padmanabhan}.} \bibinfo{year}{2011}\natexlab{}.
\newblock \showarticletitle{SCENE: a scalable two-stage personalized news
  recommendation system}. In \bibinfo{booktitle}{\emph{Proceedings of the 34th
  international ACM SIGIR conference on Research and development in Information
  Retrieval}}. ACM, \bibinfo{pages}{125--134}.
\newblock


\bibitem[\protect\citeauthoryear{Li, Zheng, Yang, and Li}{Li
  et~al\mbox{.}}{2014}]%
        {li2014modeling}
\bibfield{author}{\bibinfo{person}{Lei Li}, \bibinfo{person}{Li Zheng},
  \bibinfo{person}{Fan Yang}, {and} \bibinfo{person}{Tao Li}.}
  \bibinfo{year}{2014}\natexlab{}.
\newblock \showarticletitle{Modeling and broadening temporal user interest in
  personalized news recommendation}.
\newblock \bibinfo{journal}{\emph{Expert Systems with Applications}}
  \bibinfo{volume}{41}, \bibinfo{number}{7} (\bibinfo{year}{2014}),
  \bibinfo{pages}{3168--3177}.
\newblock


\bibitem[\protect\citeauthoryear{Lin, Xie, Guan, Li, and Li}{Lin
  et~al\mbox{.}}{2014}]%
        {lin2014personalized}
\bibfield{author}{\bibinfo{person}{Chen Lin}, \bibinfo{person}{Runquan Xie},
  \bibinfo{person}{Xinjun Guan}, \bibinfo{person}{Lei Li}, {and}
  \bibinfo{person}{Tao Li}.} \bibinfo{year}{2014}\natexlab{}.
\newblock \showarticletitle{Personalized news recommendation via implicit
  social experts}.
\newblock \bibinfo{journal}{\emph{Information Sciences}}  \bibinfo{volume}{254}
  (\bibinfo{year}{2014}), \bibinfo{pages}{1--18}.
\newblock


\bibitem[\protect\citeauthoryear{Liu, Dolan, and Pedersen}{Liu
  et~al\mbox{.}}{2010}]%
        {liu2010personalized}
\bibfield{author}{\bibinfo{person}{Jiahui Liu}, \bibinfo{person}{Peter Dolan},
  {and} \bibinfo{person}{Elin~R{\o}nby Pedersen}.}
  \bibinfo{year}{2010}\natexlab{}.
\newblock \showarticletitle{Personalized news recommendation based on click
  behavior}. In \bibinfo{booktitle}{\emph{Proceedings of the 15th international
  conference on Intelligent user interfaces}}. ACM, \bibinfo{pages}{31--40}.
\newblock


\bibitem[\protect\citeauthoryear{Liu, Wu, Wang, Li, and Wang}{Liu
  et~al\mbox{.}}{2016}]%
        {liu2016context}
\bibfield{author}{\bibinfo{person}{Qiang Liu}, \bibinfo{person}{Shu Wu},
  \bibinfo{person}{Diyi Wang}, \bibinfo{person}{Zhaokang Li}, {and}
  \bibinfo{person}{Liang Wang}.} \bibinfo{year}{2016}\natexlab{}.
\newblock \showarticletitle{Context-aware sequential recommendation}. In
  \bibinfo{booktitle}{\emph{Data Mining (ICDM), 2016 IEEE 16th International
  Conference on}}. IEEE, \bibinfo{pages}{1053--1058}.
\newblock


\bibitem[\protect\citeauthoryear{Ludewig and Jannach}{Ludewig and
  Jannach}{2018}]%
        {ludewig2018evaluation}
\bibfield{author}{\bibinfo{person}{Malte Ludewig} {and}
  \bibinfo{person}{Dietmar Jannach}.} \bibinfo{year}{2018}\natexlab{}.
\newblock \showarticletitle{Evaluation of Session-based Recommendation
  Algorithms}.
\newblock \bibinfo{journal}{\emph{arXiv preprint arXiv:1803.09587}}
  (\bibinfo{year}{2018}).
\newblock


\bibitem[\protect\citeauthoryear{McAuley, Targett, Shi, and Van
  Den~Hengel}{McAuley et~al\mbox{.}}{2015}]%
        {mcauley2015image}
\bibfield{author}{\bibinfo{person}{Julian McAuley},
  \bibinfo{person}{Christopher Targett}, \bibinfo{person}{Qinfeng Shi}, {and}
  \bibinfo{person}{Anton Van Den~Hengel}.} \bibinfo{year}{2015}\natexlab{}.
\newblock \showarticletitle{Image-based recommendations on styles and
  substitutes}. In \bibinfo{booktitle}{\emph{Proceedings of the 38th
  International ACM SIGIR Conference on Research and Development in Information
  Retrieval}}. ACM, \bibinfo{pages}{43--52}.
\newblock


\bibitem[\protect\citeauthoryear{Mikolov, Sutskever, Chen, Corrado, and
  Dean}{Mikolov et~al\mbox{.}}{2013}]%
        {mikolov2013}
\bibfield{author}{\bibinfo{person}{Tomas Mikolov}, \bibinfo{person}{Ilya
  Sutskever}, \bibinfo{person}{Kai Chen}, \bibinfo{person}{Greg~S Corrado},
  {and} \bibinfo{person}{Jeff Dean}.} \bibinfo{year}{2013}\natexlab{}.
\newblock \showarticletitle{Distributed representations of words and phrases
  and their compositionality}. In \bibinfo{booktitle}{\emph{Advances in neural
  information processing systems}}. \bibinfo{pages}{3111--3119}.
\newblock


\bibitem[\protect\citeauthoryear{Mishra and Reddy}{Mishra and Reddy}{2016}]%
        {mishra2016bottom}
\bibfield{author}{\bibinfo{person}{Sonu~K Mishra} {and} \bibinfo{person}{Manoj
  Reddy}.} \bibinfo{year}{2016}\natexlab{}.
\newblock \showarticletitle{A bottom-up approach to job recommendation system}.
  In \bibinfo{booktitle}{\emph{Proceedings of the Recommender Systems
  Challenge}}. ACM, \bibinfo{pages}{4}.
\newblock


\bibitem[\protect\citeauthoryear{Mohallick and {\"O}zg{\"o}bek}{Mohallick and
  {\"O}zg{\"o}bek}{2017}]%
        {mohallick2017exploring}
\bibfield{author}{\bibinfo{person}{Itishree Mohallick} {and}
  \bibinfo{person}{{\"O}zlem {\"O}zg{\"o}bek}.}
  \bibinfo{year}{2017}\natexlab{}.
\newblock \showarticletitle{Exploring privacy concerns in news recommender
  systems}. In \bibinfo{booktitle}{\emph{Proceedings of the International
  Conference on Web Intelligence}}. ACM, \bibinfo{pages}{1054--1061}.
\newblock


\bibitem[\protect\citeauthoryear{Moreira}{Moreira}{2017}]%
        {moreira2018chameleon}
\bibfield{author}{\bibinfo{person}{Gabriel de Souza~Pereira Moreira}.}
  \bibinfo{year}{2017}\natexlab{}.
\newblock \showarticletitle{CHAMELEON - A Deep Learning Meta-Architecture for
  News Recommender Systems}. In \bibinfo{booktitle}{\emph{RecSys'18 Doctoral
  Symposium, Proceedings of the Twelfth ACM Conference on Recommender
  Systems}}. ACM, \bibinfo{pages}{421--425}.
\newblock


\bibitem[\protect\citeauthoryear{Pennington, Socher, and Manning}{Pennington
  et~al\mbox{.}}{2014}]%
        {pennington2014glove}
\bibfield{author}{\bibinfo{person}{Jeffrey Pennington},
  \bibinfo{person}{Richard Socher}, {and} \bibinfo{person}{Christopher
  Manning}.} \bibinfo{year}{2014}\natexlab{}.
\newblock \showarticletitle{Glove: Global vectors for word representation}. In
  \bibinfo{booktitle}{\emph{Proceedings of the 2014 conference on empirical
  methods in natural language processing (EMNLP)}}.
  \bibinfo{pages}{1532--1543}.
\newblock


\bibitem[\protect\citeauthoryear{Quadrana, Karatzoglou, Hidasi, and
  Cremonesi}{Quadrana et~al\mbox{.}}{2017}]%
        {quadrana2017personalizing}
\bibfield{author}{\bibinfo{person}{Massimo Quadrana},
  \bibinfo{person}{Alexandros Karatzoglou}, \bibinfo{person}{Bal{\'a}zs
  Hidasi}, {and} \bibinfo{person}{Paolo Cremonesi}.}
  \bibinfo{year}{2017}\natexlab{}.
\newblock \showarticletitle{Personalizing Session-based Recommendations with
  Hierarchical Recurrent Neural Networks}. In
  \bibinfo{booktitle}{\emph{Proceedings of the 11th ACM Conference on
  Recommender Systems}}.
\newblock


\bibitem[\protect\citeauthoryear{R}{R}{2015}]%
        {graff2015}
\bibfield{author}{\bibinfo{person}{Graff. R}.} \bibinfo{year}{2015}\natexlab{}.
\newblock \bibinfo{title}{How the Washington Post used data and natural
  language processing to get people to read more news}.
\newblock
  \bibinfo{howpublished}{\url{https://knightlab.northwestern.edu/2015/06/03/how-the-washington-posts-clavis-tool-helps-to-make-news-personal/}}.
    (\bibinfo{date}{June} \bibinfo{year}{2015}).
\newblock


\bibitem[\protect\citeauthoryear{Rao, Jia, Feng, and Zhao}{Rao
  et~al\mbox{.}}{2013}]%
        {rao2013personalized}
\bibfield{author}{\bibinfo{person}{Junyang Rao}, \bibinfo{person}{Aixia Jia},
  \bibinfo{person}{Yansong Feng}, {and} \bibinfo{person}{Dongyan Zhao}.}
  \bibinfo{year}{2013}\natexlab{}.
\newblock \showarticletitle{Personalized news recommendation using ontologies
  harvested from the web}. In \bibinfo{booktitle}{\emph{International
  Conference on Web-Age Information Management}}. Springer,
  \bibinfo{pages}{781--787}.
\newblock


\bibitem[\protect\citeauthoryear{Ren and Feng}{Ren and Feng}{2013}]%
        {ren2013concert}
\bibfield{author}{\bibinfo{person}{Hongda Ren} {and} \bibinfo{person}{Wei
  Feng}.} \bibinfo{year}{2013}\natexlab{}.
\newblock \showarticletitle{Concert: A concept-centric web news recommendation
  system}. In \bibinfo{booktitle}{\emph{International Conference on Web-Age
  Information Management}}. Springer, \bibinfo{pages}{796--798}.
\newblock


\bibitem[\protect\citeauthoryear{Ruocco, Skrede, and Langseth}{Ruocco
  et~al\mbox{.}}{2017}]%
        {ruocco2017inter}
\bibfield{author}{\bibinfo{person}{Massimiliano Ruocco}, \bibinfo{person}{Ole
  Steinar~Lillest{\o}l Skrede}, {and} \bibinfo{person}{Helge Langseth}.}
  \bibinfo{year}{2017}\natexlab{}.
\newblock \showarticletitle{Inter-Session Modeling for Session-Based
  Recommendation}. In \bibinfo{booktitle}{\emph{Proceedings of the 2nd Workshop
  on Deep Learning for Recommender Systems}}. ACM, \bibinfo{pages}{24--31}.
\newblock


\bibitem[\protect\citeauthoryear{Salakhutdinov, Mnih, and Hinton}{Salakhutdinov
  et~al\mbox{.}}{2007}]%
        {salakhutdinov2007}
\bibfield{author}{\bibinfo{person}{Ruslan Salakhutdinov},
  \bibinfo{person}{Andriy Mnih}, {and} \bibinfo{person}{Geoffrey Hinton}.}
  \bibinfo{year}{2007}\natexlab{}.
\newblock \showarticletitle{Restricted Boltzmann machines for collaborative
  filtering}. In \bibinfo{booktitle}{\emph{Proceedings of the 24th
  international conference on Machine learning}}. ACM,
  \bibinfo{pages}{791--798}.
\newblock


\bibitem[\protect\citeauthoryear{Smirnova and Vasile}{Smirnova and
  Vasile}{2017}]%
        {smirnova2017contextual}
\bibfield{author}{\bibinfo{person}{Elena Smirnova} {and}
  \bibinfo{person}{Flavian Vasile}.} \bibinfo{year}{2017}\natexlab{}.
\newblock \showarticletitle{Contextual Sequence Modeling for Recommendation
  with Recurrent Neural Networks}. In \bibinfo{booktitle}{\emph{Proceedings of
  the 11th ACM Conference on Recommender Systems}}.
\newblock


\bibitem[\protect\citeauthoryear{Song, Elkahky, and He}{Song
  et~al\mbox{.}}{2016}]%
        {song2016multi}
\bibfield{author}{\bibinfo{person}{Yang Song}, \bibinfo{person}{Ali~Mamdouh
  Elkahky}, {and} \bibinfo{person}{Xiaodong He}.}
  \bibinfo{year}{2016}\natexlab{}.
\newblock \showarticletitle{Multi-rate deep learning for temporal
  recommendation}. In \bibinfo{booktitle}{\emph{Proceedings of the 39th
  International ACM SIGIR conference on Research and Development in Information
  Retrieval}}. ACM, \bibinfo{pages}{909--912}.
\newblock


\bibitem[\protect\citeauthoryear{Spangher}{Spangher}{2015}]%
        {spangher2015}
\bibfield{author}{\bibinfo{person}{A. Spangher}.}
  \bibinfo{year}{2015}\natexlab{}.
\newblock \bibinfo{title}{Building the Next New York Times Recommendation
  Engine}.
\newblock
  \bibinfo{howpublished}{\url{https://open.blogs.nytimes.com/2015/08/11/building-the-next-new-york-times-recommendation-engine/}}.
    (\bibinfo{date}{Aug} \bibinfo{year}{2015}).
\newblock


\bibitem[\protect\citeauthoryear{Trevisiol, Aiello, Schifanella, and
  Jaimes}{Trevisiol et~al\mbox{.}}{2014a}]%
        {Trevisiol2014}
\bibfield{author}{\bibinfo{person}{Michele Trevisiol},
  \bibinfo{person}{Luca~Maria Aiello}, \bibinfo{person}{Rossano Schifanella},
  {and} \bibinfo{person}{Alejandro Jaimes}.} \bibinfo{year}{2014}\natexlab{a}.
\newblock \showarticletitle{Cold-start news recommendation with
  domain-dependent browse graph}. In \bibinfo{booktitle}{\emph{Proceedings of
  the ACM Recommender System conference, RecSys}}, Vol.~\bibinfo{volume}{14}.
\newblock


\bibitem[\protect\citeauthoryear{Trevisiol, Aiello, Schifanella, and
  Jaimes}{Trevisiol et~al\mbox{.}}{2014b}]%
        {trevisiol2014cold}
\bibfield{author}{\bibinfo{person}{Michele Trevisiol},
  \bibinfo{person}{Luca~Maria Aiello}, \bibinfo{person}{Rossano Schifanella},
  {and} \bibinfo{person}{Alejandro Jaimes}.} \bibinfo{year}{2014}\natexlab{b}.
\newblock \showarticletitle{Cold-start news recommendation with
  domain-dependent browse graph}. In \bibinfo{booktitle}{\emph{Proceedings of
  the 8th ACM Conference on Recommender systems}}. ACM,
  \bibinfo{pages}{81--88}.
\newblock


\bibitem[\protect\citeauthoryear{Van~den Oord, Dieleman, and Schrauwen}{Van~den
  Oord et~al\mbox{.}}{2013}]%
        {van2013deep}
\bibfield{author}{\bibinfo{person}{Aaron Van~den Oord}, \bibinfo{person}{Sander
  Dieleman}, {and} \bibinfo{person}{Benjamin Schrauwen}.}
  \bibinfo{year}{2013}\natexlab{}.
\newblock \showarticletitle{Deep content-based music recommendation}. In
  \bibinfo{booktitle}{\emph{Advances in neural information processing
  systems}}. \bibinfo{pages}{2643--2651}.
\newblock


\bibitem[\protect\citeauthoryear{Vincent, Larochelle, Bengio, and
  Manzagol}{Vincent et~al\mbox{.}}{2008}]%
        {vincent2008extracting}
\bibfield{author}{\bibinfo{person}{Pascal Vincent}, \bibinfo{person}{Hugo
  Larochelle}, \bibinfo{person}{Yoshua Bengio}, {and}
  \bibinfo{person}{Pierre-Antoine Manzagol}.} \bibinfo{year}{2008}\natexlab{}.
\newblock \showarticletitle{Extracting and composing robust features with
  denoising autoencoders}. In \bibinfo{booktitle}{\emph{Proceedings of the 25th
  international conference on Machine learning}}. ACM,
  \bibinfo{pages}{1096--1103}.
\newblock


\bibitem[\protect\citeauthoryear{Wang and Blei}{Wang and Blei}{2011}]%
        {wang2011collaborative}
\bibfield{author}{\bibinfo{person}{Chong Wang} {and} \bibinfo{person}{David~M
  Blei}.} \bibinfo{year}{2011}\natexlab{}.
\newblock \showarticletitle{Collaborative topic modeling for recommending
  scientific articles}. In \bibinfo{booktitle}{\emph{Proceedings of the 17th
  ACM SIGKDD international conference on Knowledge discovery and data mining}}.
  ACM, \bibinfo{pages}{448--456}.
\newblock


\bibitem[\protect\citeauthoryear{Wang, Wang, and Yeung}{Wang
  et~al\mbox{.}}{2015}]%
        {wang2015}
\bibfield{author}{\bibinfo{person}{Hao Wang}, \bibinfo{person}{Naiyan Wang},
  {and} \bibinfo{person}{Dit-Yan Yeung}.} \bibinfo{year}{2015}\natexlab{}.
\newblock \showarticletitle{Collaborative deep learning for recommender
  systems}. In \bibinfo{booktitle}{\emph{Proceedings of the 21th ACM SIGKDD
  International Conference on Knowledge Discovery and Data Mining}}. ACM,
  \bibinfo{pages}{1235--1244}.
\newblock


\bibitem[\protect\citeauthoryear{Wang and Wang}{Wang and Wang}{2014}]%
        {wang2014}
\bibfield{author}{\bibinfo{person}{Xinxi Wang} {and} \bibinfo{person}{Ye
  Wang}.} \bibinfo{year}{2014}\natexlab{}.
\newblock \showarticletitle{Improving content-based and hybrid music
  recommendation using deep learning}. In \bibinfo{booktitle}{\emph{Proceedings
  of the 22nd ACM international conference on Multimedia}}. ACM,
  \bibinfo{pages}{627--636}.
\newblock


\bibitem[\protect\citeauthoryear{Wu, Wang, Liu, and Liu}{Wu
  et~al\mbox{.}}{2016b}]%
        {wu2016recurrent}
\bibfield{author}{\bibinfo{person}{Caihua Wu}, \bibinfo{person}{Junwei Wang},
  \bibinfo{person}{Juntao Liu}, {and} \bibinfo{person}{Wenyu Liu}.}
  \bibinfo{year}{2016}\natexlab{b}.
\newblock \showarticletitle{Recurrent neural network based recommendation for
  time heterogeneous feedback}.
\newblock \bibinfo{journal}{\emph{Knowledge-Based Systems}}
  \bibinfo{volume}{109} (\bibinfo{year}{2016}), \bibinfo{pages}{90--103}.
\newblock


\bibitem[\protect\citeauthoryear{Wu, DuBois, Zheng, and Ester}{Wu
  et~al\mbox{.}}{2016a}]%
        {wu2016}
\bibfield{author}{\bibinfo{person}{Yao Wu}, \bibinfo{person}{Christopher
  DuBois}, \bibinfo{person}{Alice~X Zheng}, {and} \bibinfo{person}{Martin
  Ester}.} \bibinfo{year}{2016}\natexlab{a}.
\newblock \showarticletitle{Collaborative denoising auto-encoders for top-n
  recommender systems}. In \bibinfo{booktitle}{\emph{Proceedings of the Ninth
  ACM International Conference on Web Search and Data Mining}}. ACM,
  \bibinfo{pages}{153--162}.
\newblock


\end{thebibliography}

\end{document}